\documentclass[11pt]{article} 

\def\tr{\:{\rm tr}\:}
\def\sgn{\, {\rm sgn}\,}
   
\def\pr{\prime}
\def\eps{\epsilon}	\def\la{\lambda}	   
\def\tht{\theta}	\def\om{\omega}

\newcommand{\tl}[1]{\tilde{#1}}
\newcommand{\pdr}{\partial}
\newcommand{\un}[1]{\underline{#1}}
\newcommand{\half}{\frac{1}{2}}
\newcommand{\ov}[1]{\frac{1}{#1}}
\newcommand{\PP}{{\scriptscriptstyle ++}}
\newcommand{\Pm}{{\scriptscriptstyle +-}}
\newcommand{\MP}{{\scriptscriptstyle -+}}
\newcommand{\MM}{{\scriptscriptstyle --}}

\newcommand{\beq}{\begin{equation}}
\newcommand{\eeq}{\end{equation}}
\newcommand{\beqs}{\begin{eqnarray}}
\newcommand{\eeqs}{\end{eqnarray}}

\mathchardef\mhyphen="2D 
\newcommand{\finteq}{\; \mhyphen\mkern-15.1mu\int} 
\newcommand{\finttxt}{\, \mhyphen\mkern-12.0mu\int} 

\usepackage{titlesec}
\titleformat{\section}{\normalsize\bfseries}{\thesection}{1em}{}
\titleformat{\subsection}{\small\bfseries}{\thesubsection}{1em}{}
\titleformat{\subsubsection}{\small\bfseries}{\thesubsubsection}{1em}{}

\usepackage{txfonts} 
\usepackage{esint} 
\usepackage{hyperref}
\usepackage{graphicx}
\usepackage{subfigure} 

\begin{document}


\begin{titlepage}

\title{\normalsize 
\hfill {\tt DCPT-10/17; arXiv:1005.4942 [hep-th]
} \\ 
\vskip 0mm 
\Large\bf On lightest baryon and its excitations in large-$N$ $1+1$ dimensional QCD}

\author{Govind S. Krishnaswami}
\date{\normalsize Department of Mathematical Sciences \& Centre for Particle Theory, \\ Durham University, Science Site, South Road, Durham, DH1 3LE, UK \vspace{.2in}\\
Chennai Mathematical Institute, \\
Padur PO, Siruseri 603103, India.
\smallskip \\ e-mail: \tt govind.krishnaswami@durham.ac.uk \\ July 31, 2010
}

\maketitle

\begin{quotation} \noindent {\large\bf Abstract} \medskip \\

We study baryons in multicolour QCD$_{1+1}$ via Rajeev's gauge-invariant reformulation as a non-linear classical theory of a bilocal meson field constrained to lie on a Grassmannian. It is known to reproduce 't Hooft's meson spectrum via small oscillations around the vacuum, while baryons arise as topological solitons. The lightest baryon has zero mass per colour in the chiral limit; we find its form factor. It moves at the speed of light through a family of massless states. To model excitations of this baryon, we linearize equations for motion in the tangent space to the Grassmannian, parameterized by a bilocal field $U$. A redundancy in $U$ is removed and an approximation is made in lieu of a consistency condition on $U$. The baryon spectrum is given by an eigenvalue problem for a hermitian singular integral operator on such tangent vectors. Excited baryons are like bound states of the lightest one with a meson. Using a rank-1 ansatz for $U$ in a variational formulation, we estimate the mass and form factor of the first excitation.

\end{quotation}





\vspace{1cm}

\centerline{\em J. Phys. A: Math. Theor. 43 (2010) 395401}

\thispagestyle{empty}

\end{titlepage}


\footnotesize

\tableofcontents


\normalsize

\section{Introduction and summary}
\label{s:intro}

An interesting problem of theoretical physics is to find the spectrum and structure of hadrons \cite{pdg} from QCD. Besides direct numerical approaches, we are far from formulating this problem in 3+1d, though there has been recent progress in the 2+1d pure gauge model \cite{Nair-karabali,rajeev-massgap}. In 1+1d, 't Hooft obtained\cite{thooft-planar-2d-mesons} an equation for masses and form factors of mesons in the multicolour $N \to \infty$ limit of  QCD. There are an infinite number of them with squared-masses growing linearly ${\cal M}_n^2 \sim \tl g^2 n$. The coupling $\tl g^2 = g_{YM}^2 N$ has dimensions of mass$^2$, so the model is UV finite. Our aim is to do the same for the spectrum of baryons in QCD$_{1+1}$. Baryons are more subtle than mesons, it hasn't been possible to extend 't Hooft's summation of planar diagrams to find the baryon spectrum\cite{wittenN}. A way forward was shown in Rajeev's formulation \cite{rajeev-2dqhd,rajeev-str-fns-from-qcd} of QCD$^{N=\infty}_{1+1}$ as a non-linear classical theory of quark bilinears ({\em meson fields}) on a curved phase space. As $N \to \infty$, gauge-invariant bilinears $M$ have small fluctuations and satisfy non-linear {\em classical} equations, though $\hbar =1$. Some non-linearities are due to a constraint on $M$ encoding Pauli exclusion. 't Hooft's meson equation was rederived by considering oscillations around the vacuum, with masses of ${\cal O} (N^0)$. But the model also has large departures from the vacuum, describing baryons with masses of ${\cal O}(N)$. They live on a disconnected component of phase space, an infinite Grassmannian with components labelled by baryon number. This formulation gave a qualitative picture \cite{ipm,solitonparton} of the baryon (as a soliton of the meson field and as a bound state of quarks) and estimates for the mass and form factor of the lightest baryon\cite{gsk-thesis}. The latter was in reasonable agreement with numerical calculations \cite{hornbostel}. They were also used to model the $x_B$-dependence of the nucleon structure function $F_3(x_B, Q^2)$ measured in deep inelastic scattering \cite{ipm,bogoliubov-anti-quark}. 

Here we wish to derive an equation for the spectrum of small oscillations around the lightest baryon, to describe excited baryons or baryon-meson bound states. For simplicity we consider $1$ quark flavour, so these correspond to the nucleon resonances $P_{11} , D_{13}, S_{11}, D_{15}$ etc \cite{pdg}. There may also exist heavier baryonic extrema of energy, analogs of $\Delta, \Lambda$. Their investigation  and oscillations around them is postponed. Oscillations near a baryon are harder to study than near the vacuum (\S \ref{s:tHooft-eqn-small-osc}). To begin, we need the precise baryon ground state (g.s.). The form factor of the lightest baryon is well-described by a single valence quark wavefunction $\psi$. In the chiral limit of massless quarks, the g.s. is exactly determined via $\psi$. We find $\psi$ exactly and establish that the lightest baryon has zero mass/colour (\S \ref{s:lightest-baryon}), like the lightest meson \cite{thooft-planar-2d-mesons}. The soliton has a size $\sim P^{-1}$ where $P$ is the mean null-momentum/colour of the baryon. Being massless, the baryon moves at the speed of light traversing a $1$-parameter family of even parity massless states. The probability of finding a valence quark with positive null-momentum between $[p,p+ {\rm d}p]$ in a baryon is $P^{-1} \exp{(-p/P)} \:  {\rm d}p$. Away from the chiral limit, the g.s. of the baryon is massive, containing sea and antiquarks\cite{bogoliubov-anti-quark}. Here we work in the simpler chiral limit. It is possible to derive\cite{ipm} this soliton picture as a Hartree-Fock approximation to $N$ quarks interacting via a linear potential, with wavefunction antisymmetric in colour but symmetric otherwise. This is a way of seeing that the baryon is a fermion and that $N$ is an integer.

As in 't Hooft's work, excitations around the translation-invariant Dirac vacuum were described by Rajeev\cite{rajeev-2dqhd} using a meson `wavefunction' $\tl \chi(\xi)$. Around a non-translation-invariant baryon, we need the $N \to \infty$ limit of a bilocal field $M(x,y) \sim q^{a \dag}(x) q_a(y)/N$\footnote{We work in a gauge where the parallel transport from $x$ to $y$ is the identity.}. The vacuum is $M =0$ while the baryon g.s. is $M_o = -2 \psi \psi^\dag$. A complication arises from a quadratic constraint $(\eps + M)^2 =1$, the `quark density matrix' must be a projection operator, up to normal ordering. We ensure it is satisfied at all times (\S \ref{s:eom-preserves-constraint}), and when making approximations (\S \ref{s:lin-eom-preserves-lin-constraint}). Pleasantly, when linearized around the baryon $M = M_o + V$, the constraint $[\eps + M_o,V]_+=0$ encodes an `orthogonality' of ground and excited states crucial for consistency of the linearized equations (\S \ref{s:H2--zero-consistency-condition}). This condition generalizes the vanishing dot product of radius $\eps + M_o$ and tangent $V$ to a sphere. Roughly, $V$ is a meson and $M_o +V$ is a meson-baryon pair. If $M_o =0$ we return to mesonic oscillations around the Dirac vacuum. Due to translation invariance around $M_o=0$, the bilocal field $\tl V(p,q) \sim \tl \chi(\xi)$ could be taken to depend only on $\xi = p/(p-q)$ and not on the `total momentum' $p-q$. This simplification is absent near the baryon (\S \ref{s:failure-of-meson-type-ansatz}). So in \S \ref{s:constraint-linearized} we solve the constraint $[\eps + M_o , V]_+=0$ via another bilocal field $V = {\rm i}[\eps + M_o,U]$. But there is a {\em gauge freedom} under $U \to U + U_g$ where $[\eps + M_o, U_g]=0$. We gauge-fix the redundancy (\S \ref{s:gauge-freedom-in-U}) by writing $U$ in terms of a vector $u$ and another bilocal field $U^\Pm$ a fourth the size of $U$. Roughly, $u$ is a correction to the valence quarks $\psi$, due to the excitation. $U^\Pm$ has the corresponding data on sea/antiquarks in the excited baryon. The gauge fixing conditions $\psi^\dag u =0$ and $\psi^\dag U^\Pm = 0$ are interpreted as orthogonality of ground and excited states. But the naively linearized equations don't preserve these conditions! The gauge freedom at each time-step is used to derive linearized equations respecting the gauge conditions (\S \ref{s:linear-eom-for-U}). Though the equations for $U^\Pm$ and $u$ are linear, we weren't able to find oscillatory solutions by separation of variables. For, they couple $u, U^\Pm$ {\em and} their adjoints, like a Schrodinger equation where the hamiltonian depends on the wavefunction and its conjugate! So in \S \ref{s:u-equal-0-approx} we put $u=0$, allowing us to separate variables and find oscillatory solutions, at the cost of a consistency condition on $U^\Pm$ (\ref{e:consistency-cond-on-U-PM}). Regarding $V$ as a meson, we expect it contains a quark-antiquark sea but no valence quarks $u$. This motivates the $u=0$ ansatz.

We are left with an eigenvalue problem $\hat K(U) = \om U$ (\ref{e:eigenvalue-problem-for-U-PM}) for the form factor $U^\Pm$. We show that the linearized hamiltonian $\hat K$ is hermitian using the gauge condition and the ansatz $u=0$. In the chiral limit, the mass$^2$ of excited baryons are ${\cal M}^2 = 2 \om P$, where $P$ is the lightest baryon's momentum. But the eigenvalue problem for $\hat K$ is quite non-trivial. It is a singular integral operator on a `physical subspace' of hermitian operators. This space of physical states $U^\Pm$ consists of Hilbert-Schmidt operators subject to the gauge and consistency conditions (\S \ref{a:solve-consistency-condition}). The eigenvalue problem for the baryon spectrum follows from a variational energy $\cal E$. In \S\ref{s:rank-1-ansatz-for-Upm} we suggest a rank-$1$ variational ansatz $U^\Pm = \phi \eta^\dag$. $\phi, \eta$ are the sea/antiquark wavefunctions of the excited baryon. The kinetic terms in $\cal E$ differ from the naive ones due to linearisation around a time-dependent g.s. The potential energy is a sum of Coulomb energy (attraction between anti and sea-quarks) and exchange energy (between sea-partons and `background' valence quarks $\psi$). In \S \ref{s:estimate-mass-shape-1st-excitation} we obtain a crude estimate for the mass and form factor of the first excited baryon by minimising $\cal E$ in a parameter controlling the decay of the sea quark wavefunction. But our estimate for the mass of the 1st excited baryon $.3 \tl g N$ isn't expected to be accurate\footnote{From 't Hooft's work\cite{thooft-planar-2d-mesons} the mass of the 1st excited meson in the chiral limit is about $1.4 \tl g$.} or an upper bound, as we imposed the gauge-fixing condition but not the consistency condition from the ansatz $u=0$. In \S\ref{a:solve-consistency-condition} we try to solve this consistency condition. A more careful treatment will hopefully give a quantitative understanding of the baryon spectrum. 


\subsection{Summary of Classical Hadrondynamics}
\label{s:review-chd}

We begin by recalling Rajeev's reformulation \cite{rajeev-2dqhd} of QCD$^{N = \infty}_{1+1}$ as a classical theory of meson fields. In null coordinates $x = x^1, t = x^0 - x^1$ we specify initial values on the null line $t=0$. Energy $E = p_t = p_0$ and null-momentum $p = p_x = p_0 + p_1$ obey\footnote{Under a Lorentz boost of rapidity $\tht$, $t \to t {\rm e}^\tht$ and $x \to {\rm e}^{-\tht} x - t \sinh{\tht}$.} $m^2 = 2Ep - p^2$. In the gauge $A_x = A_0 + A_1 = 0$, one component of quarks and the gluon $A_1$ are eliminated.
For quarks of one flavour and $N$ colours $a,b$, the action of SU$(N)$ QCD$_{1+1}$ represents fermions $\chi_a$ interacting via a linear potential
	\beq
	S = \int {\rm d}t \, {\rm d}x \, \chi^{\dag a} \left[- {\rm i} \pdr_t - {\textstyle \half \left(p + \frac{m^2}{p}\right)} \right] \chi_a - {\textstyle \frac{g^2}{4 N}} \int {\rm d}t {\rm d}x {\rm d}y \, \chi^{\dag a}(y) \chi_b(y) |x-y| \chi^{\dag b}(x) \chi_a(x).
	\label{e:action-QCD-after-elim}
 	\eeq
$\hat M(x,y) \! =\! - \frac{2}{N} \! : \! \chi^{\dag a}(x) \chi_a(y) \! :$ with $x,y$ null-separated, defines a gauge-invariant bilocal field. Normal-ordering is with respect to the Dirac vacuum. $E$ and $p$ have the same sign, so -ve momentum states are filled in the vacuum and we split the 1-particle Hilbert space ${\cal H} = L^2(\mathbf{R}) = {\cal H}_- \oplus {\cal H}_+$ into $\mp$ momentum states\footnote{Our convention for Fourier transforms is $\psi(x) = \int [{\rm d}p] \:  {\rm e}^{{\rm i}px} \tl \psi(p)$ where $2\pi \: [{\rm d}p] = {\rm d}p$. 
}. Canonical anti-commutation relations (CAR) for $\chi, \chi^\dag$ from (\ref{e:action-QCD-after-elim}) imply commutation relations for $\hat M$, with fluctuations of order $1/N$. As $N \to \infty$, $\hat M$ tends to a classical field $M$, the integral kernel\footnote{In Fourier space, $\tl M(p,q) = \int {\rm d}x\, {\rm d}y\,  {\rm e}^{- {\rm i} (px-qy)} M_{xy}$. We write $\tl M_{pq}$ for $\tl M(p,q)$ and $M_{xy}$ for $M(x,y)$.} of a hermitian operator on $\cal H$. Poisson brackets of $M$ are given by 
    \beq
    ( {\rm i} / 2) \: \{M(x,y),M(z,u)\} = \delta(z-y) \Phi(x,u) - \delta(x-u) \Phi(z,y).
    \label{e:pb-of-M}
    \eeq
$\Phi = \eps + M$ where $\eps$ is the Hilbert transform kernel $\tl \eps(p,q) = 2\pi \delta(p-q) \sgn p$, or $\eps(x,y) = \frac{ {\rm i} }{\pi} {\cal P} \left({x-y} \right)^{-1}$. The CAR imply a constraint as $N \to \infty$, $\Phi^2 = I$: eigenvalues of $\Phi$ are $-1$ (singly-occupied) or $1$ (unoccupied). $\Phi = \eps$ is the vacuum. Thus the {\em phase space} is a Grassmannian \cite{rajeev-2dqhd}
	\beq
	{\rm Gr}_1 = \left\{ M \: : \:  M^\dag = M, \; (\eps + M)^2=I, \: \tr |[\eps,M]|^2 < \infty  \right\},
	\label{e:Gr-1}
	\eeq
the symplectic leaf of $\Phi = \eps$ under the coadjoint action of a restricted unitary group \cite{rajeev-2dqhd}. The coadjoint orbit formula for Poisson brackets of linear functions of $M$, $f_u = -\half \tr u M$ is
	\beq
	\{ f_u, f_v \} \: = \: {\textstyle \frac{ {\rm i} }{2}} \tr [u,v] \Phi \: =\: f_{- {\rm i} [u,v]} + {\textstyle \frac{ {\rm i} }{2}} \tr[u,v] \eps.
	\label{e:pb-of-linear-fns}
	\eeq
The connected components of Gr$_1$ are labelled by an integer $B = -\half \tr M$ (\S\ref{a:convergence-conditions-inner-product}), quark number per colour, or baryon number. An analogue of parity is $\mathbf{P} \tl M_{pq}(t) = \tl M_{qp}(-t)$ or $\mathbf{P} M_{xy}(t) = M_{-x,-y}^*(-t)$. E.g. static real symmetric $\tl M$ are even and imaginary antisymmetric $\tl M$ are odd. From (\ref{e:action-QCD-after-elim}), the energy/colour is a parity-invariant quadratic function on Gr$_1$, 
	\beq
	E(M) = -{\textstyle \half} \int {\textstyle \half \left(p + \frac{\mu^2}{p} \right)} \tl M(p,p) [{\rm d}p] + {\textstyle \frac{\tilde g^2}{16}} \int |M(x,y)|^2 {|x-y|} \: {\rm d}x \, {\rm d}y.
        \label{e:energy-of-CHD}
    \eeq
The current quark mass $m$ is renormalized $\mu^2 \! =\! m^2 - \frac{\tl g^2}{\pi}$ while reordering quark bilinears. The kinetic-energy $T = -\half \tr h M$ is expressed in terms of the dispersion kernel 
	\beq
	\tl h(p,q) = 2\pi \delta(p-q) h(p) \;\;\; \textrm{where} \quad
	2 \: h(p) = {\textstyle p + \mu^2 p^{-1} }.
	\label{e:dispersion-relation-hofp}
	\eeq
Define a positive `interaction operator' on hermitian matrices $\hat G : M \mapsto G(M) \equiv G_M$ with kernel $\hat G(M)_{xy} = {\textstyle \half} M_{xy}{|x-y|}$ (\S \ref{a:evaluate-G-of-Mo}). Then the potential energy is
	\beq {\textstyle 
	U = {\textstyle \frac{\tl g^2}{8}} \tr M \hat G(M) \: = \: 
	{\textstyle \frac{\tl g^2}{16}} \int {\rm d}x\, {\rm d}y\, |M(x,y)|^2 \: |x-y| \geq 0. }
	\label{e:potn-egy-U-intermsof-GofM}
	\eeq
In Fourier space\footnote{This uses $v(x) = \half |x| = - \finttxt \frac{[{\rm d}r]}{r^2} {\rm e}^{-{\rm i}rx}$ got by solving $v^{\pr \pr}(x) = \delta(x)$ with $v(0)=v'(0)=0$. We used the definition of finite part integrals (\S\ref{a:finite-part-integrals}) to put $\finttxt_{-\infty}^\infty \frac{[{\rm d}r]}{r} = 0$ and $\finttxt_{-\infty}^\infty \frac{[{\rm d}r]}{r^2} = 0$.} $\tl G(M)_{pq} = -\finttxt {\textstyle \frac{[{\rm d}r]}{r^2}} \tl M_{p+r,q+r}$. We also associate to $M$ a constant of motion (\S \ref{a:cons-of-momentum}), its mean momentum per colour $P_M$. Under a boost, $P \to {\rm e}^\tht P$, $E \to {\rm e}^{-\tht} E + p \sinh \tht$.
    \beq
        P_M = -{\textstyle \half} \tr {\tt p} M 
        = -{\textstyle \half} \int p \tilde M(p,p) [{\rm d}p] \;\; {\rm where} \;\;    {\tt p}(p,q) = 2\pi \delta(p-q) p.
        \label{e:momentum-of-CHD}
    \eeq
The squared-mass/colour ${\cal M}^2 \, = \, 2 E P - P^2$ is a Lorentz-invariant constant of motion. Hamilton's equations of motion (eom) are the initial value problem (IVP)
    \beq
    {\textstyle \frac{ {\rm i} }{2} \frac{{\rm d} M}{{\rm d}t}} = {\textstyle \frac{ {\rm i} }{2}} \{E(M), M\} = [E'(M),\eps + M].
	\label{e:hamiltons-eqns}
    \eeq
The P.B. is expressed via the commutator using the variational derivative of energy, which is inhomogeneous linear in $M$, $E'= T' + U' = - h/2 + ({\tl g^2 / 4}) \hat G(M)$. Its matrix elements are
    \beq
    E'(M)_{pq} = - \pi \delta(p-q) h(p) + {\textstyle \frac{\tilde g^2}{4}} \tl G(M)_{pq} \quad {\rm where} \;\; U'(M)_{xy} \equiv {\textstyle \frac{\delta U(M)}{\delta M_{yx}}} = {\textstyle \frac{\tl g^2}{4} \frac{|x-y| }{ 2}} M_{xy}.
    \label{e:E-prime-of-M}
    \eeq
\subsection{Preservation of quadratic constraint under time evolution}
\label{s:eom-preserves-constraint}

We check that (\ref{e:hamiltons-eqns}) preserve the constraint $\Phi^2 = I$. Define the constraint matrix $C(t) = \Phi^2 - I$ and let $C(0)=0$. We have an autonomous system of 1st order non-linear ODEs 
	\beq
	\pdr_t C = \pdr_t (\eps + M)^2 = [\eps + M, \pdr_t M]_+ = -2 {\rm i} [\Phi, [E', \Phi]]_+ = -2 {\rm i} [E'(M(t)), \Phi^2(t)].
	\label{e:ivp-for-quadratic-constraint}
	\eeq
Under suitable hypotheses, it should have a unique solution\footnote{Rhs is a cubic function of $\Phi$. Picard iteration should establish that the solution to (\ref{e:ivp-for-quadratic-constraint}) exists and is unique. We may need technical hypotheses (besides $\tr |[\eps, M(0)]|^2 < \infty$ \S \ref{a:convergence-conditions-inner-product}) on $\Phi(0)$, to ensure observables (e.g. energy) remain finite.} given $C(0)$. Now consider the guess $C_{g}(t) \equiv 0$. It obeys (\ref{e:ivp-for-quadratic-constraint}) as both sides vanish: $\pdr_t C_g(t)= 0$ and $-2 {\rm i} [E', \Phi^2(t)] = -2 {\rm i} [E', I] = 0$. Thus $C_{g}(t) \equiv 0$ is the solution: constraint is always satisfied.

\section{Ground state in \texorpdfstring{$B=0$}{} meson sector}
\label{s:gs-in-meson-sector}

In the non-interacting case $\tl g =0$, $M=0$ is a static solution since the eom are
	\beq
	{\textstyle \frac{ {\rm i} }{2}} \dot M_{pq} \:= \: 
	{\textstyle \ov{4}} M_{pq} \left[ q-p + m^2 \left( {\textstyle \ov{q} - \ov{p}} \right) \right]
	\quad {\rm when} \quad \tl g \to 0.
	\label{e:eom-non-interacting}
	\eeq
Rhs$\equiv0$ iff $M \!= \!0$, so it is the {\em only} static solution if $\tl g =0$. Even with interactions, $M=0$ is static: $\pdr_t M = \{ E(M), M \} =0$ at $M=0$ (\ref{e:energy-of-CHD}). But even at $M=0$, $E'(0) = - \pi \delta(p-q) h(p)$ does not vanish! Does the gradient of energy vanish at $M=0$? Yes. To see why, first note that $M=0$ is a static solution as $E'(0)$ and $\eps$ are diagonal in momentum space. By (\ref{e:hamiltons-eqns})
	\beq
	\pdr_t M = -2{\rm i}[E'(M),\eps + M]\vert_{{M=0}} = - 2 {\rm i} [E'(0),\eps] =0.
	\eeq
$E'(M)=0$ is sufficient, but not necessary for a static solution. 
$-2i[E'(M),\Phi]$ is the symplectic gradient of energy at $M$. The contraction of the exterior derivative of energy with the Poisson bivector field produces the Hamiltonian vector field. So the (symplectic) gradient of energy does vanish at $M=0$. The state $M=0$ has zero mass $\cal M$ and qualifies as a g.s.

\section{Small Oscillations about vacuum and 't Hooft's meson equation}
\label{s:tHooft-eqn-small-osc}

We recall the equation for mesons \cite{thooft-planar-2d-mesons,rajeev-2dqhd} by considering small oscillations about the vacuum. Let $V$ be a tangent vector at the translation-invariant $M=0$. Constraint $\Phi^2 = I$ becomes\footnote{$V^\Pm_{pq}: {\cal H}_- \to {\cal H}_+$ has entries with $p>0>q$, $(V^\Pm)^\dag = V^\MP$. We separate matrix rows with $|$.} $[\eps, V]_+ = 0$ or
	\beq
	\tl V_{pq} \; (\sgn p + \sgn q) =0
	\,\;\; \Rightarrow \;\;\, \tl V
	= (0, \tl V^\MP \, | \, \tl V^\Pm , 0).
	\label{e:constraint-for-mesons}
	\eeq
$\tl V_{pp}=0$, so $V$ has zero mean momentum $P_V$ (\ref{e:momentum-of-CHD}). But the generator $P_t = p-q$ of translations $M_{xy} \to M_{x+a,y+a}$, $\tl M_{pq} \to {\rm e}^{ {\rm i} (p-q)a} \tl M_{pq}$ may be regarded as total momentum. So we pick independent variables $P_t$ and $\xi = p/P_t$. We write $\tl V^\Pm = \, \tl \chi(P_t, \xi, t)$. Hermiticity implies\footnote{$P_t \geq 0$ in the $+-$ block while $P_t \leq 0$ in the $-+$ block, but $\xi \in [0,1]$ always.}  
	\beq
	\tl V^\MP(p,q,t) = \tl \chi(P_t,\xi,t) \quad {\rm with} \quad
	\tl \chi^*(P_t,\xi,t) = \tl \chi(-P_t, 1-\xi,t).
	\label{e:hermiticity-cond-on-V-for-mesons}
	\eeq
$\xi$ is the quark momentum fraction. For small oscillations about $M=0$ of energy $\om = p_0$ we put
	\beq
	\tl V^\Pm_{pq}(t) = \tl \chi(P_t,\xi) {\rm e}^{ {\rm i} \om t} \quad {\rm and} \quad
	\tl V^\MP_{pq}(t) = \tl \chi(P_t,\xi) {\rm e}^{- {\rm i} \om t}
	\;\; {\rm for}  \;\; \om \in \mathbf{R}.
	\label{e:time-dependence-for-meson-ansatz}
	\eeq
Parity acts as ${\bf P} \tl \chi = \tl \chi^*$. The simplest $\tl \chi$ obeying (\ref{e:hermiticity-cond-on-V-for-mesons}) are independent of $P_t$ with $\tl \chi^*(\xi) = \tl \chi(1-\xi)$. So even parity states are real with $\tl \chi(\xi) = \tl \chi(1-\xi)$ and odd parity ones imaginary with $\tl \chi(\xi) = -\tl \chi(1-\xi)$. The norm (\ref{a:convergence-conditions-inner-product}) on $V$ implies the $L^2$ norm on $\tl \chi(\xi)$ upto a divergent constant. The linearized eom are
	\beqs
	{\textstyle \frac{ {\rm i} }{2}} \dot{V} \!\!\!\! &=& \!\!\! [E'(V), \Phi] =  \left[T' + {\textstyle \frac{1}{4}} {\tl g^2} G(V) ,\, \Phi \right] 
	 = [T', V] + {\textstyle \frac{\tl g^2}{4}}
		[G(V), \eps] + {\cal O}(V^2), \cr
	{\textstyle \frac{ {\rm i} }{2}} \pdr_t \tl V_{pq} \!\!\!\! &=& \!\! -{\textstyle \half} \left\{h(p) - h(q) \right\} \tl V_{pq}
		- {\textstyle \frac{\tl g^2}{4}}  \: (\sgn q - \sgn p) \finteq \frac{[{\rm d}s]}{s^2}
	     \tl V_{p+s,q+s}.
	\label{e:deriving-tHooft-eqn-1}
	\eeqs
Put $\eta' = s/P_t$ to get an eigenvalue problem for $\om$. It is rewritten as 't Hooft's equation for the squared-masses ${\cal M}^2 = 2 \om P_t - P_t^2$ with quarks of equal mass \cite{thooft-planar-2d-mesons} ($\mu^2 = m^2 - \frac{\tl g^2 }{ \pi}$, $\eta = \xi + \eta'$). For instance, with $\mu^2  =0$, the eigenstates alternate in parity $\tl \chi_n(\xi) \approx {\rm i}^{n-1} \sin(n\pi \xi)$ with mass$^2$ ${\cal M}^2_n \approx n \pi \tl g^2$.
	\beqs
	- {\textstyle \frac{\om}{2}} \tl \chi(\xi) \!\!\! &=& \!\!\!
	- {\textstyle \ov{4}} \left[P_t + \frac{\mu^2}{\xi P_t} + \frac{\mu^2}{P_t - \xi P_t} \right] \tl \chi(\xi) + {\textstyle \frac{\tl g^2}{2}} \finteq \frac{\tl \chi(\xi + \eta') }{ \eta'^2 P_t} [{\rm d} \eta'], 
	\cr
	&& \!\!\!  {\cal M}^2 \tl \chi(\xi) \, = \, \left(\frac{\mu^2}{\xi} + \frac{\mu^2}{1-\xi} \right) \tl \chi(\xi) - {\textstyle \frac{\tl g^2}{\pi}} \finteq_0^1 \frac{\tl \chi(\eta)}{(\xi - \eta)^2} {\rm d}\eta.
	\label{e:tHooft-meson-eqn}
	\eeqs

\section{Ground state of baryon}
\label{s:lightest-baryon}

The trajectories $M_o(t)$ of least mass on the $B=1$ component are the baryonic g.s, they depend on $m,\tl g$. The chiral limit is $m \to 0$ holding $\tl g$ fixed, $\nu = {m^2 / \tl g^2} \to 0$. Regarding QCD$_{1+1}$ as an approximation to QCD$_{3+1}$ on integrating out directions $\perp$ to hadron propagation, $\tl g^{-1} \sim {\cal O}($transverse hadron size$)$. So the chiral limit should describe u/d quarks, that are much lighter than the size of hadrons. But it is hard to find the g.s. from the non-linear eom (\ref{e:hamiltons-eqns}). Inspired by valence partons, we found that the g.s. is approximately  rank-1 \cite{gsk-thesis,ipm,solitonparton}. $M = -2 \psi \psi^\dag$ lies on the $B=1$ component if $\tl \psi$ is a +ve momentum ($\eps \psi = \psi$) unit vector. 
We guessed that a minimum mass +parity state is\footnote{$P = -\tr {\tt p} M/2$ (\ref{e:momentum-of-CHD}) is baryon momentum/colour, it fixes the frame. A rescaling of $p$ \& $P$ is a boost.}
	\beq \textstyle
	\tl {M_0}_{pq} = -\frac{4\pi}{P} {\rm e}^{-\frac{p+q}{2 P}} \: \tht(p) \tht(q), 
	\:\:\:
	\tl \psi_0(p) = \sqrt{\frac{2\pi }{ P}} {\rm e}^{\frac{-p}{2P}} \tht(p), 
	\:\:\:
	\psi_0(x) = \ov{\sqrt{2\pi P}} \left[ \ov{(2P)^{-1} - {\rm i}x} \right].
	\label{e:exp-ansatz-for-baryon-gs}
	\eeq
In \S\ref{s:mass-of-exp-ansatz} we show (\ref{e:exp-ansatz-for-baryon-gs}) has zero mass as $\nu \to 0$. In \S\ref{s:degeneracy-time-evol-baryon-gs} we show it is one of a family of  degenerate massless states connected by time evolution. $M_t$ is thus a baryon g.s., 
	\beq {\textstyle
	\tl {M_t}_{pq} = \tl {M_0}_{pq} {\rm e}^{ {\rm i} (p-q) t/2}, 
	\:\:\:\:
	\tl \psi_t(p) = {\rm e}^{ipt/2} \tl \psi_0(p),
	\:\:\:\:
	\psi_t(x) = {\textstyle \ov{\sqrt{2 \pi P}} \left[ \ov{2P} - {\rm i} \left(x + \frac{t}{2} \right) \right]^{-1} }.}
	\label{e:exp-ansatz-1-parameter-family}
	\eeq
$p$-$q$ is {\em not} a constant, unlike near the translation-invariant $M=0$ (\S \ref{s:tHooft-eqn-small-osc}). Since $M_{xx} \sim [ \left(x+t/2 \right)^2 + (2P)^{-2} ]^{-1}$, the baryon is localized at $x = -t/2$ at time $t$ and has a size $\sim 1/P$. As $x=x^1, t= x^0 - x^1$, the massless baryon travels at the speed of light\footnote{So though the null line $t=0$ is not a Cauchy surface, the baryon trajectory intersects it.} along $x^1 = - x^0$. The probability of finding a valence quark of momentum $p$ in the baryon is $-\half \tl M(p,p)$\footnote{The off-forward pdfs of deeply virtual Compton scattering \cite{dvcs} depend on off-diagonal entries of $M$.}. So the degeneracy and time-dependence are consequences of relativity: a massless soliton can't be at rest. Time-dependent vacua are unusual\footnote{They are forbidden in elementary QM: energy eigenstates must have simple-harmonic time dependence. But if the g.s. of a QFT describes a massless particle whose number is conserved, it can't be static. Classical evolution allows more possibilities. A near-example is of a pair of like charges. The {\em unattainable} g.s. is for them to be at rest infinitely apart. A state of finite separation can't be static: repelling charges accelerate.}. Continuously connected static vacua (states of neutral equilibrium) are more common, e.g. the g.s. of a ball on a horizontal plane. There are time-dependent states of arbitrarily small energy $>0$, where the ball adiabatically rolls between vacua. What is remarkable about $M_t$ is that there is no `additional kinetic energy of rolling between vacua', due to the masslessness of the quarks. But this massless baryon is special to the chiral limit. Away from $m=0$, the g.s. of the baryon is roughly $M = -2 \psi \psi^\dag$, with $\tl \psi(p) \propto p^a {\rm e}^{-p/2P} \tht(p)$, $a \approx \sqrt{3 \nu /\pi}$ and ${\cal M}^2 \approx \tl g^2 \sqrt{\pi \nu / 3}$ for small $\nu$ \cite{gsk-thesis}.

\subsection{Mass of the separable exponential ansatz}
\label{s:mass-of-exp-ansatz}

To find the mass of (\ref{e:exp-ansatz-for-baryon-gs}), we split energy (\ref{e:energy-of-CHD}) as $2 E = P + {m^2} {\rm KE} + {\tl g^2} ({\rm SE} + {\rm PE})$ where $\tl g^2 {\rm SE}/2$ is a self-energy. In terms of $\nu = m^2/\tl g^2$, the mass$^2$ $2 E P - P^2$ is given by
	\beqs
		\label{e:mass-squared-defn}
	{\cal M}^2 &=&  \tl g^2 \, P\,  (\nu {\rm KE} + {\rm SE} + {\rm PE}) 
	\, \stackrel{m \to 0}{\longrightarrow} \, \tl g^2 P ({\rm SE + PE}), 
		\quad {\rm where} \\
	{\rm PE} &=& {\textstyle \ov{4} \int {\rm d}x\:  {\rm d}y\:  |M_{xy}|^2 \frac{|x-y|}{2},\: \: \: \: 
	{\rm SE} = \ov{2\pi} \int \tl M_{pp} \frac{[{\rm d}p]}{p},\: \: \: \: 
	{\rm KE} = -\half \int \tl M_{pp} \frac{[{\rm d}p]}{p}.}
	\nonumber
	\eeqs
For $M = -2 \psi \psi^\dag$, PE $=\int {\rm d}x \, |\psi|^2 V(x)$ where $V = \half \int {\rm d}y |\psi(y)|^2 |x-y|$ obeys $V^{\pr \pr} = |\psi|^2$,
	\beqs
	{\textstyle
	V(0) = \half \int {\rm d}y\:  |\psi(y)|^2 \: |y|  {\rm \: \: \: \:  and}\: \: \: \:  V'(0)} \!\!\!\!\! &= {\textstyle - \half \int_{-\infty}^\infty {\rm d}y \: |\psi(y)|^2 \sgn{y}.}
	\label{e:bc-for-self-cons-potn-in-posn-space}
	\\
	{\textstyle {\rm Thus,} \;\;
	{\rm PE} = \int [{\rm d}p]\:  \tl \psi(p) \int [{\rm d}r]\,  \tl \psi^*(p+r)\,  \tl V(r)
\: \: \, {\rm where}} & {\textstyle
	\tl V = {\textstyle \frac{-1}{r^2}} \int [{\rm d}q] \, \tl \psi(q) \, \tl \psi^*(q-r).}
	\label{e:PE-in-terms-of-V-mom-space}
	\eeqs
Here, $|\psi_o(y)|^2 = \ov{2\pi P} [(2P)^{-2} + y^2]^{-1}$ is even, so $V'(0)=0$ and $\tl V(r)$ is real and even. But $V(0)$, SE and PE are $\log$-divergent. Yet, we will show that SE + PE$=0$, regarded as a limit of regulated integrals\footnote{To bypass the regularisation, we can set up rules for manipulating these integrals based on the answers we get via the regularized calculations. From (\ref{e:PE-in-terms-of-V-mom-space}) the potential energy is
	\beq
	2 \pi P \, {\rm PE} 
	= - \int_0^\infty {\rm d}q  \,  {\rm e}^{-q} \finteq_{-q}^\infty \frac{{\rm d}s }{ s^2} {\rm e}^{-\frac{s + |s| }{ 2}}
	= - \int_0^\infty {\rm d}q  \,  {\rm e}^{-q} \left[\finteq_{-q}^0 \frac{{\rm d}s}{s^2} + \finteq_0^\infty \frac{{\rm d}s}{s^2} {\rm e}^{-s} \right] 
	= \finteq_0^\infty \frac{{\rm d}s}{s} \,  {\rm e}^{-s} - \finteq_0^\infty \frac{{\rm d}s}{s^2} \,  {\rm e}^{-s} .
	\eeq
These terms are equal by integration by parts if we ignore the boundary term. So for PE + SE $= 0$, we must define
	\beq {\textstyle
	(\pi P) \:  {\rm SE} = - \finttxt_0^\infty  {\rm d}s \: s^{-1} \: {\rm e}^{-s}
	\equiv \finttxt_0^\infty {\rm d}s \: s^{-2} \: {\rm e}^{-s}
	\quad {\rm or} \quad s^{-1} {\rm e}^{-s} \vert_{s=0} \equiv 0.}
	\eeq
}
	\beqs 
	\tl V(r) \!\!\! &=& \!\!\! {\textstyle - \ov{r^2 P} {\rm e}^{r/2P} 
		\int_{\max(0,r)}^\infty {\rm e}^{-q/P} {\rm d}q  \: = \:  - \ov{r^2} 
		\exp\left( - \frac{|r|}{2P} \right)}; 
	\nonumber
	\\
	{\rm SE}  \!\!\! &=& \!\!\! {\textstyle \ov{2\pi} \int \tl M_{pp} \frac{[{\rm d}p]}{p}  	\: = \: \frac{-1}{\pi P} \int_0^\infty \frac{{\rm e}^{-q}}{q} { {\rm d}q}}
	 \:\:\: {\rm and} \:\:\: 
	{\rm PE} = \:  {\textstyle \ov{4 \pi^2 P} \int {\rm d}x {\rm d}y \frac{|x-y|}{(1 + x^2)(1 + y^2)}}.
	\eeqs

\subsubsection{Regularized/Variational estimation of baryon ground state}
\label{s:regularized-variational-ansatz-baryon-gs}

Let us use an IR regulator to ensure PE, SE are finite. Let $\tl \psi(p) \sim p^a {\rm e}^{-p} \tht(p)$ so that $\tl \psi$ is continuous at $p=0$ if $a>0$. For $a=0$, this reduces to our ansatz $\psi_o$ in the frame with $2P = 1$. We regard this as an ansatz for minimising ${{\cal M}^2}$ (\ref{e:mass-squared-defn}). We show ${{\cal M}^2}$ vanishes as $a \to 0$ if $\nu =0$. Let
	\beqs 
	\tl \psi_a(p) = {\textstyle \frac{2^{1+a} \sqrt{\pi} }{ \sqrt{\Gamma{(1+2a)}} } p^{a}{\rm e}^{-p} \:  \tht(p)}, && \!\!
    \psi_a(x) = {\textstyle \frac{\sqrt{\Gamma{(1+2a)}} }{ 2^a \Gamma{(\half + a)}} \ov{(1-ix)^{1+a}}} 
     \:\:\:\:\: {\rm for \:\:  which}, 
     \nonumber
     \\
	P(a) = { \textstyle \int p |\tl \psi_p|^2 [{\rm d}p]= \half + a}, \!\! && \!\!\!\!
    {\rm KE} = { \textstyle \int |\tl \psi_p|^2 \frac{[{\rm d}p]}{2p}= \frac{1}{a}}, \; \;
    {\rm SE} = {\textstyle - \int |\tl \psi_p|^2 \frac{[{\rm d}p]}{\pi p} = \frac{-1}{\pi a}}.
    \eeqs
Integrating and imposing the initial condition $V_a(0) = \Gamma(a) \, / \, [2 \sqrt{\pi} \Gamma(a + 1/2) ]$,
	\beq \textstyle
	V'_a(x) =	\frac{x \Gamma (a+1) \,  _2F_1\left(\half,a+1;
	 \frac{3}{2};-x^2\right)}{\sqrt{\pi } \Gamma (a+\half)}, 
    \:\:\:\: 
	V_a(x) = \frac{\Gamma (a) \left(2 a x^2 \,  _2F_1\left(\half , a+1;\frac{
   3}{2};-x^2\right) + \left(x^2+1\right)^{-a}\right)}{2 \sqrt{\pi }
   \Gamma \left(a + \half \right)}.
	\eeq
Note that $_2F_1\left(\half , a+1; \frac{3}{2};-x^2\right) \propto x^{-1}$ for large $x$ and $a > 0$, so $V_a(x) \propto |x|$ as $|x| \to \infty$. However, we couldn't do the final integral to get ${\rm PE} = \int {\rm d}x \:  V_a \:  |\psi_a|^2$. It converges for $a > 0$ as $V_a \:  |\psi_a|^2 \sim |x|^{-1-2a}$ as $|x| \to \infty$. Upon integrating for some simple values of $a$ we find that SE + PE $\to 0$ as $a \to 0$.
\begin{table}[htdp] \vspace{-.2cm} {\footnotesize
\caption{\label{t:pe-plus-se-vs-a}\footnotesize Though ${\rm PE} \approx 6.9 \times 10^{-7} + .3183/a + 1.046 \:  a - 4.3 \:  a^2$, grows as $a \to 0$, SE+PE $\propto {\cal M}^2$ decreases. 
} \vspace{-.2cm}
\begin{center}
\begin{tabular}{@{}c|cccccccc@{}}
\hline
$a$ & $ 0.1 $ & $ 0.01 $ & $ 0.005 $ & $ .00333 $ & $ 0.00167 $ & $ 0.00125 $ & $ 0.001 $ & $ 0.0001 $ \\
PE & 3.255 & 31.84 & 63.67 & 95.5 & 190.99 & 254.65 & 318.3 & 3183 \\
SE + PE & .07209 & .01003 & .005123 & .00344 & .001733 & .001302 & .001043 & .000105 \\
\hline
\end{tabular}
\end{center} \vspace{-.2cm}}
\end{table}
We fit a series\footnote{It is tempting to Laurent expand the integrand in $a$ and integrate term by term. But this doesn't work as the operations of integration and Laurent expansion do not commute:
	\beqs
	V_a(x) = (2\pi a)^{-1} + (2\pi)^{-1} \left(2x \arctan{x} - \log \left\{ (1+ x^2) / 4 \right\} \right) + \cdots  \: = \:  V_{-1} a^{-1} + V_0 + V_1 a + \cdots \cr
	|\psi_a(x)|^2 = \left( \pi(1+x^2) \right)^{-1} \left[1 - a \log \left\{(1 + x^2) / 4 \right\} + \cdots \right]  \: = \:  |\psi|^2_0 + |\psi|^2_1 a + \cdots
	\eeqs
Integrating term by term, the 1st converges $\ov{a} \int V_{-1} |\psi|^2_0 = \ov{2 \pi a}$, but to half the numerical value.
	\beq
	{\rm PE} = \int {\rm d}x V(x) |\psi(x)|^2 \stackrel{?}{=} \ov{a} \int V_{-1} |\psi|^2_0
	+ \int (V_{-1} |\psi|^2_1 + V_0 |\psi|^2_0) + a \int (V_{-1} |\psi|^2_2 + V_0 |\psi|^2_1 + V_1 |\psi|^2_0) + \cdots
 	\eeq
The $a^0$ term diverges $V_0 |\psi|^2_0 \sim |x|^{-1}$, $\int V_0 |\psi|^2_1$ also diverges: expanding in $a$ destroys convergence of the integral!} to the calculated PE (Table \ref{t:pe-plus-se-vs-a}) for several $a \in [10^{-2}, 10^{-4}]$. It is plausible that the coefficient of $1/a$ is {\em exactly} $1/\pi \simeq .3183$ and cancels SE$=-1/{\pi a}$ and moreover, that PE + SE vanishes at $a = 0$. Encouraged by this, we calculated PE$(a)$ using Mathematica for several round values of $a^{-1}$. There was a pattern and we conjectured (\ref{e:conj-formula-for-PEofa}), which was confirmed for 100's of $a$'s. We are confident PE + SE vanishes as $a \to 0$. So $M_o = -2 \psi_0 \psi_0^\dag$ (\ref{e:exp-ansatz-for-baryon-gs}) is a massless baryon in the chiral limit: $\lim_{m \to 0} {\cal M}^2 / \tl g^2 =0$. From (\ref{e:mass-squared-defn}), its energy/colour is $E_o = P/2$ where $P$ is its momentum/colour.
	\beq \textstyle
	{\rm PE}(a) \stackrel{?}{=} \frac{\Gamma(a) \Gamma(\half + 2a) }{ 4^a \Gamma(\half + a)^3} = \ov{\pi a} + \frac{\pi}{3} a - \frac{12 \:  \zeta(3) }{ \pi} a^2 + {\cal O}(a^3).
	\label{e:conj-formula-for-PEofa}
	\eeq

\subsection{Degeneracy and time-dependence of massless baryon states in chiral limit}
\label{s:degeneracy-time-evol-baryon-gs}

We generalize the massless baryon $M_o$ (\ref{e:exp-ansatz-for-baryon-gs}) to a family $M_t$ (\ref{e:exp-ansatz-1-parameter-family}). $M_t$ clearly lie on the $B=1$ component. Further, $P(t) = -\half \tr {\tt p} M_0(t) = P(0) =P$, KE$(t) =$ KE$(0)$, SE$(t) =$ SE$(0)$ and by going to position space, PE$(t) = $PE$(0)$. So $M_t$ is massless (\ref{e:mass-squared-defn}) like $M_0$ (\ref{e:exp-ansatz-for-baryon-gs}). We found $M_t$ by time-evolving $M_0$ in the chiral limit, so $t$ is time. 
$M_0$ evolves according to $ \dot M = \{E(M), M \}$ (\ref{e:hamiltons-eqns}):
	\beq
	{\textstyle \frac{ {\rm i} }{2}} \dot M_{pq} 
	= {\textstyle \half} \tl M_{pq} {[h(q) - h(p)]} - {\textstyle \frac{\tl g^2}{4}} G(M)_{pq} [\sgn p - \sgn q] - {\textstyle \frac{\tl g^2}{4}} [ M, G_M]_{pq}.
	\label{e:eom-for-M}
	\eeq
We must show $M_t$ obeys the eom $ \frac{ {\rm i} }{2} \dot M_{pq} =\ov{4} \left( q-p \right) M_{pq} + \frac{\tl g^2}{4} Z(M)_{pq}$ where
	\beq
	Z(M)_{pq} =
	{\textstyle \ov{\pi}} \left({\textstyle \ov{p} - \ov{q}} \right) \tl M_{pq} - G(M)_{pq} \{\sgn p - \sgn q \} - [M, G_M]_{pq}.
	\label{e:eom-chiral-limit}
	\eeq
In \S\ref{a:Z-equals0} we show that $Z(M(t)) \equiv 0$ for all $t$, so the interactions cancel out! Now
	\beq
	M(t)_{pq} = M_{pq}(0) {\rm e}^{\frac{{\rm i}}{2} (p-q) t}  \:\: \Rightarrow  \:\: 
	{\textstyle \frac{{\rm i}}{2}} \dot M_{pq}(t) = {\textstyle \ov{4}} (q-p) M_{pq}(t).
	\eeq
So $M_0$ evolves to $M_t$ with energy $P/2$, describing a baryon moving at the speed of light.

\section{Small oscillations about lightest baryon}
\label{s:small-osc}

\subsection{Linearisation and solution of constraint on perturbation \texorpdfstring{$V$}{}}
\label{s:constraint-linearized}

Suppose $M_o(t)$ is the g.s. for $B=1$ with momentum $P_o = -\half \tr {\sf p} M_o$. Write $M = M_o + V$ where $V$ is a small perturbation tangent to Gr$_1$ at $M_o(t)$. Then $V^\dag = V$ and $\tr V=0$. $V$ is a meson and $M_o + V$ a baryon-meson pair. What are the masses and form factors of excited baryons? The constraint $\Phi^2 = 1$ linearizes to $[\eps + M_o, V]_+ = 0$. This generalizes $v \cdot \phi + \phi \cdot v =0$ for tangent vectors to $S^2$. Now,
	\beq
 	\eps + M_o = \pmatrix{ -1 & 0 \cr 0  & 1 + M_o^\PP}
	 \: \Rightarrow \: 
	[\eps + M_o, V]_+ = \pmatrix{ -2 V^\MM & V^\MP M_o^\PP \cr
	M_o^\PP V^\Pm & 2 V^\PP + [M_o^\PP, V^\PP]_+ }= 0.
	\label{e:constraint-linear-blockform}
	\eeq
In particular, $V^\MM =0$. Roughly, $V^\MP M_o^\PP =0$ expresses orthogonality of ground and excited states. (\ref{e:constraint-linear-blockform}) is solved\footnote{We haven't shown this is the {\em most general} solution of (\ref{e:constraint-linear-blockform}). By analogy with the sphere, we suspect that the anti-commutant of $\Phi$ is the image of the adjoint action $ {\rm i} \: {\rm ad}_{\Phi}$ on hermitian matrices.} by introducing a hermitian matrix $U$, a `potential' for $V$. $V = {\rm i}[\Phi_o, U]$ is automatically traceless, hermitian and anti-commutes with $\Phi_o$. This generalizes $v = \phi \times u$ for a tangent vector to $\phi \cdot \phi =1$. 
Motivated by (\ref{e:exp-ansatz-1-parameter-family}), let $M_o(t) = -2 \psi \psi^\dag$ be a separable baryon state, then 
	\beq
	V = {\rm i} \pmatrix{ 0 & \!\!\!\! -U^\MP(2+M_o^\PP) 
	\cr (2+M_o^\PP)U^\Pm & [M_o^\PP,U^\PP] }
	  =  2 {\rm i} \pmatrix{ 0 & \!\!\!\!\! -U^\MP (1 - \psi \psi^\dag) \cr
	(1 - \psi \psi^\dag) U^\Pm & [U^\PP, \psi \psi^\dag] }.
	\label{e:soln-of-constraint-for-V}
	\eeq
Here $1 = I^\PP$, is the identity on ${\cal H}_+$. We let $U^\MM=0$, it doesn't contribute. $U^\PP$ and $U^\Pm$ are the unknowns. Recall that for mesonic oscillations around $M=0$, the constraint implied $V^\PP=0= U^\PP$. 



\subsection{{\em Gauge fixing} freedom in choice of \texorpdfstring{$U$}{} for fixed \texorpdfstring{$V = {\rm i} [\Phi_o, U]$}{}}
\label{s:gauge-freedom-in-U}

Our solution $V = {\rm i} [\Phi_o, U]$ to the constraint (\ref{e:constraint-linear-blockform}) is unchanged under $U \mapsto U + U_g$ if $[U_g, \Phi_o]=0$. This generalizes the fact that if $\phi \times u = v$ is tangent to $S^2_{\phi \cdot \phi =1}$ at $\phi$, then so is $\phi \times (u + u_g)$ for any $u_g$ parallel to $\phi$. We eliminate this redundancy by imposing a gauge condition picking out one member from each equivalence class $U \sim U + U_g$. A convenient condition can be used to kill some entries of $U$. To understand the extent of the gauge freedom, we first find the commutant $\{ \Phi_o \}^\pr$, i.e. the pure gauge matrices $[\Phi_o, U_g] =0$. For $M_o = -2 \psi \psi^\dag$ with $\eps \psi = \psi$, and $\psi^\dag \psi =1$, this becomes
	\beq
	{\rm (i)} \:\:\:  [P_\psi, U_g^\PP]  \: = \: 0 
	\;\;\; {\rm and} \;\;\;
	{\rm (ii)} \:\:\:  P_\psi U_g^\Pm  \: = \:  U_g^\Pm.
	\label{e:pure-gauge-conditions}
	\eeq
$P_\psi = \psi \psi^\dag$ projects to ${\rm span}(\psi)$ in ${\cal H}_+$. (i) says that $U_g^\PP \in \{ P_\psi \}'$, which we characterize by extending $\psi_0 \equiv \psi$ to an o.n. basis for ${\cal H}_+$: $\{ \psi_k \}_0^\infty$. The commutant of $P_\psi$ consists of the hermitian matrices
	\beq
	U_g^\PP = a_{00} \psi_0 \psi_0^\dag + {\textstyle \sum_{k,l \geq 1}} \: a_{kl} \psi_k \psi_l^\dag  
	= \left( a_{00}, 0 \, | \, 0 , A \right) 
	\:\:\: {\rm with} \:\: 
	a_{00} \in \mathbf{R}.
	\label{e:pure-gauge-Upp}
	\eeq
Here $A:{\rm span}_\psi^\perp \to {\rm span}_\psi^\perp$. 
To find $U_g^\Pm$ let $\{ \eta_k \}_0^\infty$ be an o.n. basis for ${\cal H}_-$ and write (\ref{e:pure-gauge-conditions})(ii) as 
	\beq
	U_g^\Pm = {\textstyle \sum_{k,l \geq 0}} \;\: u_{kl}  \:  \psi_k  \:  \eta_l^\dag = 
	P_\psi U^\Pm_g = {\textstyle \sum_{l \geq 0}} \;\: u_{0l}  \: \psi_0  \:  \eta_l^\dag.
	\eeq
The solution is $u_{kl}=0$ for $k \ne 0$ and $u_{0l}$ arbitrary. (\ref{e:pure-gauge-Upp},\ref{e:pure-gauge-Upm}) characterize the pure-gauge $U_g$.
	\beq
	U_g^\Pm  \! = \!  {\textstyle \sum_{l \geq 0}} \; u_{0l} \:  \psi_0 \:  \eta_l^\dag 
	 \: = \:   
	 \pmatrix{ u_{00} & u_{01} & \cdots & u_{0l} 
	 & \cdots \cr & & {\bf 0} & & } 
	 \:\:\:\: {\rm with}  \:\:\:  u_{0l} \:\:\: {\rm arbitrary}.
	\label{e:pure-gauge-Upm}
	\eeq
{\noindent \sf Gauge-fixing conditions:} The gauge freedom (\ref{e:pure-gauge-Upm}) is used to kill the 1st row of $U^\Pm$. 
This is equivalent to imposing $P_\psi U^\Pm =0$ or $\psi^\dag U^\Pm =0$. Similarly, the pure gauge $U^\PP_g$'s (\ref{e:pure-gauge-Upp}) can be used to kill the $00$ entry and all but the first row and column of $U^\PP$. So most of $U^\PP$ is pure gauge. Thus in the {\em mostly-zero gauge}, $U$ may be taken in the form ($\vec 0$ and $\vec u$ represent column vectors)
	\beqs
	U^\MM \!\!\!\!\! &=& \!\!\! 0,  \:\:\: 
	U^\MP = ( \vec 0  \:\:  {\bf W} ), \:\:\: 
	U^\PP = (0,\vec u^\dag \: | \: \vec u, {\bf 0} ) = u \psi^\dag + \psi u^\dag,
	 \:\:\:  {\rm where} \nonumber
	\\
	{\bf W} \!\!\!\! &:& \!\!\!\!\! {\rm span}_\psi^\perp \to {\cal H}_-;  \:\:\:\: 
	\vec u = \left( u_1 \: u_2 \: \cdots \right)^t,  \:\:\: 
	\psi_0 \perp u = {\textstyle \sum_{k \geq 1}} \; u_k \:  \psi_k \in {\cal H}_+.
	\label{e:mostly-zero-gauge}
	\eeqs
For mesonic oscillations $V^\PP \!\! = \! U^\PP \!\! = \!\! 0$ (\ref{e:constraint-for-mesons}) but around a baryon, $U^\PP$ can be taken rank-$2$. The physical degrees of freedom are encoded in a vector  $u \in {\cal H}_+$ and a matrix $(U^\MP)^\dag = U^\Pm$ in the
	\beq
	\textrm{mostly-zero gauge:} \:\:\:\:\:  \psi^\dag u = 0,  \:\:\:  \psi^\dag U^\Pm = 0  \:\:\: {\rm and} \:\:\:  U^\MM = 0.
	\label{e:mostly-zero-gauge-condition}
	\eeq
So $\psi$ is $\perp$ to the excitation $U$. E.g.\footnote{A more general example of a matrix with $\psi$ in its kernel is $U^\MP = \un U^\MP (1 - P_\psi)$ for {\em any} matrix $\un U^\MP$.} the rank-1 ansatz $U^\MP = \eta \phi^\dag$ with $\phi,\eta \in {\cal H}_\pm$ and $\phi^\dag \psi =0$. 
The g.s. time-dependence  is simple, $\tl \psi_t(p) = \tl \psi_0(p) {\rm e}^{{\rm i}pt/2}$ (\ref{e:exp-ansatz-1-parameter-family}). So if at $t=0$, $\phi_0^\dag \psi_0 =0$, then orthogonality is maintained if $\tl \phi_t(p) = \tl \phi_0(p) {\rm e}^{-{\rm i}pt/2}$. To {\em summarize}, if $U$ is picked in the gauge (\ref{e:mostly-zero-gauge-condition}), then by (\ref{e:soln-of-constraint-for-V})  
	\beq
	V = \pmatrix{ 0 & V^\MP \cr V^\Pm & V^\PP } 
	= {\rm i} \pmatrix{ 0 & -U^\MP(2+M_o) \cr
	(2+M_o) U^\Pm & [M_o,U^\PP] }  = 
	2 {\rm i} \pmatrix{ 0 & - U^\MP \cr  U^\Pm & u \psi^\dag - \psi u^\dag}.
	\label{e:mostly-zero-gauge-V-of-U}
	\eeq
Conversely, $U(V)$ is defined up to addition of a pure gauge $U_g$. Given $V$, we can find a convenient representative in the equivalence class of $U$'s that it corresponds to. In the mostly-zero gauge, upon using $u^\dag \psi =0$, we get $u = \ov{2{\rm i}} V^\PP \psi$\footnote{Since $u \perp \psi$, this is consistent only if $V^\PP \psi \perp \psi$, i.e., $\psi^\dag V^\PP \psi =0$, which is the same as the condition $\tr M_o^\PP V^\PP=0$. But this is guaranteed by the constraint (\ref{e:constraint-linear-blockform}) $2 V^\PP = [V^\PP, M_o^\PP]$ upon multiplying by $M_o^\PP$ and taking a trace.}. Thus $U^\PP = u \psi^\dag + \psi u^\dag = -\frac{{\rm i}}{4}[V^\PP, 2 \psi \psi^\dag]$. In this gauge, $U^\Pm \propto V^\Pm$. Given $V$, the most general corresponding $U$ is the sum of any $U_g \in \{\Phi_o \}^\pr$ (\ref{e:pure-gauge-Upp},\ref{e:pure-gauge-Upm}) and
	\beq
	U_{\textrm{\small mostly-zero gauge}}  \: = \:  \pmatrix{ 0 & U^\MP 
	\cr U^\Pm & U^\PP }
	 \: = \:  {\textstyle \ov{2{\rm i}}} \pmatrix{ 0 & - V^\Pm \cr 
	 V^\Pm & [V^\PP, \psi \psi^\dag] }.
	\eeq

\subsection{Linearized equations of motion for perturbation \texorpdfstring{$V$}{}}
\label{s:linearized-eom}

For $M(t) = M_o(t) + V(t)$, (\ref{e:hamiltons-eqns}) becomes $ {\rm i} \pdr_t (M_o + V) = 2 [ E'_{M_o + V} , \Phi_o + V]$. The solution describes a curve $M(t)$ on the $B=1$ component of phase space. Our g.s. is time-dependent, so this is like the effect of Jupiter on the motion of Mercury. For the nucleon, we refer to resonances created by scattering a $\pi,{\rm e}^-$ or $\nu$ off the proton. From (\ref{e:E-prime-of-M}), $E'(M_o + V) = E'(M_o) + \frac{\tl g^2}{4} G_V$, so linearising,
	\beq
	{\textstyle \frac{{\rm i}}{2}} \dot{V} = \left( - {\textstyle \frac{{\rm i}}{2}} \pdr_t M_o + [E'(M_o), \Phi_o] \right)	+ [E'(M_o) , V]
	+ {\textstyle \frac{\tl g^2}{4}} [ G_V, \Phi_o ] + {\cal O}(V^2).
	\eeq
The terms in round brackets add to zero if $M_o(t)$ satisfies the eom, as does our baryon g.s. (\ref{e:exp-ansatz-1-parameter-family}). So
	\beq
	{\textstyle \half} \dot{V} = {\rm i}[V,E'(M_o)] - {\textstyle \frac{ {\rm i} \tl g^2}{4}} 
	[G_V, \Phi_o]
	= {\rm i}[V, T'] - {\textstyle \frac{ {\rm i} \tl g^2}{4}} 
	\left\{ [G_{M_o} , V] + [G_V, \Phi_o]	\right\}.
	\label{e:time-evol-of-V-free-interact}
	\eeq
$T' = - h/2$. To see the departure from 't Hooft's meson equation write
$\dot{V} = H = H_1 + H_2$ with
	\beq
	 H_1 = {\rm i} [h,V] - {\textstyle \frac{ {\rm i} \tl g^2}{2}} 
	 [G_V, \eps]   \:\:\: {\rm and} \:\:\: 
	H_2 = -{\textstyle \frac{ {\rm i} \tl g^2}{2}} 
		\left\{ [G_{M_o}, V] + [G_V, M_o] \right\}.
	\label{e:time-evol-of-V}
	\eeq
$H_1$ is independent of $M_o$ and leads to 't Hooft's meson equation (\ref{e:deriving-tHooft-eqn-1}) if $M_o =0$.
$H_2$ has `baryon-meson' interactions leading to many complications. In blocks, the eom are
	\beq
	\pmatrix{ 0 & \!\!\!\! \dot V^\MP \cr 
	\dot V^\Pm & \!\!\!\! \dot V^\PP }
	= {\rm i} \pmatrix{ 0 & \!\!\! [h,V^\MP] \cr 
		[h,V^{\Pm}] & \!\!\! [h,V^\PP] }
	- {\textstyle \frac{ {\rm i} \tl g^2}{2} }
	\pmatrix{ [G_M,V]^\MM & \!\!\! [G_M,V]^\MP + G_V^\MP \left(2+M^\PP \right)
	\cr -{\rm h.c.} 
	& \!\!\! [G_M,V]^\PP + [G_V^\PP, M^\PP] }.
	\label{e:time-evol-of-V-blockform}
	\eeq

\subsection{Linearized time evolution preserves constraints}
\label{s:lin-eom-preserves-lin-constraint}

(\ref{e:time-evol-of-V-free-interact}) describes the motion of a point $V(t)$ in the tangent bundle of the Grassmannian restricted to the base $M_o(t)$. To establish this, we show that (\ref{e:time-evol-of-V-free-interact}) preserves hermiticity of $V$ and the linear constraint (\ref{e:constraint-linear-blockform}). If $V$ is hermitian at time $t$, then so are $G_V$, $G_{M_o}$ and $H(V)$. By (\ref{e:time-evol-of-V-free-interact}), $V(t+\delta t)$ is also hermitian. As for the linear constraint, suppose $\Phi_o(t)$ is the solution of (\ref{e:hamiltons-eqns}) about which we perturb by $V(t)$, and define a constraint function $C(t) = [\Phi_o(t), V(t)]_+$, which satisfies $C(0)=0$. Then using (\ref{e:time-evol-of-V-free-interact})
	\beq
	{\textstyle \frac{{\rm i}}{2}} \dot{C} 
	= {\textstyle \frac{{\rm i}}{2}} \left\{  [\dot{\Phi}_o, V]_+
		+ [\Phi_o, \dot{V}]_+ \right\}
	= [[E'_{M_o}, \Phi_o], V]_+ + [\Phi_o,[E'_{M_o},V]]_+
		+ {\textstyle \frac{\tl g^2}{4}} [\Phi_o, [G_V, \Phi_o]]_+.
	\eeq
To find the unique solution of this autonomous linear system of 1st order ODEs, we make the guess $C(t) \equiv 0$ which annihilates the lhs. On the rhs, the 1st two terms cancel as $[\Phi_o, V]=0$. The 3rd term vanishes as $\Phi_o^2 = I$ (\S \ref{s:eom-preserves-constraint}). So $C(t) =0$ is the unique solution and (\ref{e:time-evol-of-V-free-interact}) preserves the linear constraint. {\sf Corollary:} As both $V(t)$ and $V(t+ \delta t)$ satisfy the constraint, so does the  difference quotient $H(V(t))$. And when $H$ is split as in (\ref{e:time-evol-of-V-free-interact}), both $[E'(M_o),V]$ and $[G(V), \Phi_o]$ satisfy the linear constraint if $V$ does. But if $H$ is split as in (\ref{e:time-evol-of-V}), $H_1$ and $H_2$ {\em do not} each satisfy (\ref{e:constraint-linear-blockform}), except at $M_o=0$.

\subsection{Equation of motion in \texorpdfstring{`$--$'}{} block: orthogonality of excited states}
\label{s:H2--zero-consistency-condition}

The $--$ block of the eom (\ref{e:time-evol-of-V-blockform}) is simplest as it is non-dynamical, $[G(M_t),V(t)]^{--}=0$. This is necessary for consistency of the eom. It says $V^\MP \:  G_M^\Pm: {\cal H}_- \to {\cal H}_-$ is always hermitian:
	\beq
	G(M_t)^\MP V_t^\Pm = V_t^\MP G(M_t)^\Pm.
	\label{e:H2--equals-zero-condition}
	\eeq
Using the constraint $V^\MP M^\PP =0$ (\ref{e:constraint-linear-blockform}), we show that $V^\MP G_M^\Pm \equiv 0$! Our argument uses the exponential form of the g.s. $M_o(t)$ (\ref{e:exp-ansatz-1-parameter-family}), but there may be a more general proof. We simplify (\ref{e:H2--equals-zero-condition}) using the fact that the g.s. interaction operator (\ref{e:G-of-Mo-summary}) is always rank-1. Putting
{\setlength\arraycolsep{1pt}
	\beqs
	\tl G(M_t)^\MP_{pr} \, = \, ({2 / P}) \, {\rm e}^{-r/2P} {\rm e}^{- \frac{{\rm i}}{2}r t} 
	\!\! && \!\!
	{\rm e}^{-p/2P} \, {\rm e}^{\frac{{\rm i}}{2} p t} \, {\rm I}_2(-p) 
	 \:\:\: {\rm in} \:\:  (\ref{e:H2--equals-zero-condition}) 
	\cr
	\Rightarrow  \:\:  \int_0^\infty [{\rm d}r] \:  {\rm e}^{\frac{{\rm i}}{2}(p-r)t}  \: {\rm e}^{-\frac{p+r}{2P}} \:  I_2(-p) \:  V^\Pm_{rq} 
	&=& 
	\int_0^\infty [{\rm d}r] \:  V^\MP_{pr} \: {\rm e}^{\frac{{\rm i}}{2}(r-q)t} \:  {\rm e}^{-\frac{r+q}{2P}}  \:  I_2(-q)
	\eeqs}
for all $p,q < 0$. Dividing by $I_2 \ne 0$ (\ref{e:I2}), and using $\tl \psi_t(r) \propto \tht(r) {\rm e}^{-r (1/P - {\rm i} t)/2}$ (\ref{e:exp-ansatz-1-parameter-family}) we get
	\beq
	\frac{\int_0^\infty [{\rm d}r] \:  \tl \psi_t(r)  \:  V^\Pm_{rq}}{I_2(-q) {\rm e}^{-\frac{q}{2} (\ov{P} - {\rm i} t)}}
	 \: = \:  \frac{\int_0^\infty [{\rm d}r] \:  V^\MP_{pr}  \:  \psi_t(r) }{ I_2(-p) {\rm e}^{-\frac{p}{2} (\ov{P} - {\rm i} t)}}
	 \: = \:  c(t), \:\:\:\: 
	\forall \:\:  p,q  \: < \:  0.
 	\eeq
Lhs \& rhs depend on $q$ \& $p$, so they must be equal! $c(t) \in \mathbf{R}$ by hermiticity. So (\ref{e:H2--equals-zero-condition}) becomes
	\beq
	\int_0^\infty [{\rm d}r]  \:  \tl V^\MP_{pr} \:  \tl \psi_t(r) = c(t)  \: {\rm e}^{-\frac{p}{2} (\ov{P} - {\rm i}t)} \:  I_2(-p),  \:\:\:\:  \forall \:\:  p<0.
	\eeq
$V^\MP$ maps the g.s. to $c(t) \times$ a vector in ${\cal H_-}$. But $V$ annihilates the g.s: $V^\MP M_o^\PP =0$ (\ref{e:constraint-linear-blockform})! So $c(t) \equiv 0$,  $V^\MP G_{M_o}^\Pm =0$ and $[G_{M_o},V]^\MM \equiv 0$. It says excited states are $\perp$ to the g.s.

\subsection{Lack of translation invariance: Failure of ansatz \texorpdfstring{$V^\Pm_{pq}(t) = \tl \chi_t(\xi)$}{}}
\label{s:failure-of-meson-type-ansatz}

In the $+-$ block of the eom (\ref{e:time-evol-of-V-blockform}), let us try what worked for mesons (\S \ref{s:tHooft-eqn-small-osc}). Around the translation-invariant $M=0$ vacuum, $V^\Pm_{pq}(t) = \tl \chi_t\left(\xi, P_t \right)$ could be taken independent of $P_t = p-q$ (\ref{e:time-dependence-for-meson-ansatz}). For oscillations around a {\em non}-translation-invariant baryon $M_o$ (\ref{e:exp-ansatz-for-baryon-gs}), such an ansatz doesn't work; $P_t$ can't be regarded the momentum of $\tl V$. The {\em orthogonality} constraint $V^\MP M_o^\PP=0$ (\ref{e:constraint-linear-blockform}) is violated if $\tl \chi$ is independent of $p-q$. To see this, $V^\MP M_o^\PP=0$ is expressed using $\tl M= -2 \psi \psi^\dag$ as
	\beq
	\int_0^\infty \tl \chi(\xi,t) \:  \tl \psi_t(q) \:  {\rm d}q =0,  \:\:\:  \forall p < 0
	 \:\, \Leftrightarrow \:\,
	\int_0^1 \tl \chi(\xi,t)  \:  \tl \psi_t\left( p(1-\xi^{-1}) \right)  \: {\textstyle \frac{{\rm d}\xi }{\xi^2}} =0,  \:\:  \forall p < 0.
	\eeq
$\tl \chi_t$ must be $\perp$ to each of $f_p(\xi;t) = \tl \psi_t(p(1-1/\xi))/\xi^2$ for $p<0$ at all times $t$. E.g. at $t=0$,
	\beq
	f_p(\xi) = \xi^{-2} \psi_o\left( p(1-1/\xi) \right) \sim \xi^{-2} \exp\{-p(1-1/\xi)\}  \:\: {\rm for} \:\:\:  p < 0.
	\eeq
$f_p(\xi)$ are linearly independent +ve functions going from $f_p(0) = 0$ to $f_p(1)=1$ with maxima shifting rightwards as $0 \geq p \geq -\infty$. Plausibly, for $\tl \chi$ to be $\perp$ (in $L^2(0,1)$) to all of them requires $\tl \chi \equiv0$. So non-trivial $\tl V^\Pm_{pq}$ must depend on $p-q$. It seems prudent to work instead with the unconstrained $U$.

\subsection{Linearized evolution of unconstrained perturbation \texorpdfstring{$U$}{}}
\label{s:linear-eom-for-U}

To find the linearized evolution of $U$, we put $V = {\rm i}[\Phi_o,U]$ in (\ref{e:time-evol-of-V-free-interact})
	\beq
	 {\rm i} [\Phi_o, \dot U] = [[\Phi_o, U],h] + {\textstyle \frac{\tl g^2}{2}} [G_{[\Phi_o, U]}, \eps] +{\textstyle \frac{\tl g^2}{2}}  \left\{[G_{M_o}, [\Phi_o, U]] + [G_{[\Phi_o, U]}, M_o] \right\}.
	\eeq
Some entries of $U$ are redundant due to gauge freedom. So we derive eom in the mostly-zero gauge in terms of the vector $u$ and matrix $U^\Pm$ (\ref{e:mostly-zero-gauge}). This requires some care. The eom don't know our gauge choice, and we mustn't expect them to preserve the gauge conditions (\ref{e:mostly-zero-gauge-condition}) $\psi^\dag u =0$ and $\psi^\dag U^\Pm =0$. Using (\ref{e:mostly-zero-gauge-V-of-U}),
we begin by writing (the tentative nature of this evolution is conveyed by $\doteq$) 
	\beq
	2 {\rm i} \dot u  \: \doteq \:  V^\PP \dot \psi + \dot V^\PP \psi , 
	\qquad 2 {\rm i} \dot U^\Pm  \: \doteq \:  \dot V^\Pm.
	\label{e:eom-for-u-Upm-begin}
	\eeq
Here, $\dot{\psi}_t(p) = \half {\rm i} p \tl \psi_t(p)$, if $\psi$ is chosen as the g.s. valence quark wavefunction in the chiral limit (\ref{e:exp-ansatz-1-parameter-family}). We use the eom  (\ref{e:time-evol-of-V-blockform}) for $V$ and (\ref{e:mostly-zero-gauge-V-of-U}) to express the rhs in terms of $u,U^\Pm$. For example,
	\beqs 
	2 {\rm i} \dot u \!\!\! &\doteq& \!\!\!
	2 {\rm i} (u \psi^\dag - \psi u^\dag) \dot \psi
	+ 2 [u \psi^\dag - \psi u^\dag, h] \psi 
	\cr && \!\!\!
	- \tl g^2 \left\{[u \psi^\dag - \psi u^\dag, G_M^\PP] + G_M^\Pm U^\MP + U^\Pm G_M^\MP  \right\} \psi
	- {\rm i} \tl g^2 [\psi \psi^\dag , G_V^\PP] \psi.
	\eeqs
$G_V$ is given in \S \ref{a:G-of-V}. We regard these as equations for $(u,U^\Pm)(t+\delta t)$ given $(u,U)^\Pm(t)$ satisfying the gauge conditions (\ref{e:mostly-zero-gauge-condition}). So on the rhs we can use (\ref{e:mostly-zero-gauge-condition}) to simplify:
	\beqs
	 {\rm i} \dot u \!\! &\doteq& \!\!\!
	 {\rm i} (u \psi^\dag - \psi u^\dag) \dot \psi
	+ (u \psi^\dag - \psi u^\dag)h\psi - hu 
	\cr && \!\!
	+ {\textstyle \frac{\tl g^2}{2}} \left\{ G_M^\PP u - (u \psi^\dag - \psi u^\dag) G_M^\PP \psi 
	- U^\Pm G_M^\MP \psi
	- {\rm i} P_\psi G_V^\PP \psi + {\rm i} G_V^\PP \psi \right\},
	\cr	
	 {\rm i} \dot U^\Pm \!\!\!\! &\doteq& \!\!\! [U^\Pm,h] + {\textstyle \frac{\tl g^2}{2}} \left\{ (\psi u^\dag - u \psi^\dag) G^\Pm_M
	- U^\Pm G_M^\MM + G_M^\PP U^\Pm + {\rm i} (1 - P_\psi) G_V^\Pm \right\}.
	\label{e:eom-for-u-Upm-before-projecting}
	\eeqs
But we have a problem. This evolution does not preserve the gauge-fixing conditions:
	\beqs
	 {\rm i} {\textstyle \frac{{\rm d}}{{\rm d}t}} ( \psi^\dag u) \!\!\! &\doteq& \!\!\!
	 {\rm i} \dot \psi^\dag u
	- u^\dag h \psi - \psi^\dag h u + {\textstyle \frac{\tl g^2 }{ 2}} \left\{ \psi^\dag G_M^\PP u + u^\dag G_M^\PP \psi \right\} \ne 0,
	\cr 
	 {\rm i} {\textstyle \frac{{\rm d}}{{\rm d}t}} ( \psi^\dag U^\Pm) \!\!\! &\doteq& \!\!\!
	 {\rm i} \dot \psi^\dag U^\Pm
	- 2 \psi^\dag h U^\Pm + \tl g^2 \left\{ u^\dag G_M^\Pm + \psi^\dag G_M^\PP U^\Pm \right\} \ne 0.
	\eeqs
But at each time-step, we may add to $U(t+\delta t)$ a pure-gauge $U_g(t + \delta t)$ to bring it to the mostly-zero gauge, so that at $t+\delta t$, $\psi^\dag u=0$ and $\psi^\dag U^\Pm=0$. This corresponds to subtracting out the instantaneous projections on $\psi$ and defining a new time evolution that preserves (\ref{e:mostly-zero-gauge-condition})
	\beq
	 {\rm i} \dot u  \: := \:  {\textstyle \half} (1-P_\psi) \left(V^\PP \dot \psi + \dot V^\PP \psi \right)
	 \:\:\:\: {\rm and} \:\:\:\:  
	 {\rm i} \dot U^\Pm  \: := \:  {\textstyle \half} (1-P_\psi) \dot V^\Pm.
	\eeq
This projection involves no approximation. We use (\ref{e:mostly-zero-gauge-condition}) to simplify the rhs to get\footnote{Signs of $l,L^\Pm$ are chosen so the hamiltonian in \S \ref{s:u-equal-0-approx} is +ve. Some integrals are IR divergent if $\tl \psi(p) \propto {\rm e}^{-p/2P} \tht(p)$ is the exact chiral g.s. E.g. for the regulator of \S\ref{s:regularized-variational-ansatz-baryon-gs}, $\psi^\dag h \psi = \half (\half + a + \frac{\mu^2}{a})$. We suspect all divergences cancel in physical quantities, as for the lightest baryon. Also, most of these divergences disappear for the ansatz $u=0$ studied in \S \ref{s:u-equal-0-approx}-\ref{s:estimate-mass-shape-1st-excitation}.}.
	\beqs
	 {\rm i} \dot u \!\!\!\! &\equiv& \!\!\!\! -l = 
	 {\rm i} u \psi^\dag \dot \psi
	+ \left\{ \psi^\dag h \psi - [1 - P_\psi] h \right\} u 
	- {\textstyle \frac{\tl g^2}{2}} \left\{  \psi^\dag G_M^\PP \psi \, u 
	+ U^\Pm G_M^\MP \psi 
	- [1-P_\psi] \left( G_M^\PP u + {\rm i} G_V^\PP \psi \right) \right\}
	\cr
	 {\rm i} \dot U^\Pm \!\!\!\! &\equiv& \!\!\!\! -L^\Pm \! = 
	U^\Pm h - [1-P_\psi] h U^\Pm - {\textstyle \frac{\tl g^2}{2}} \left\{u \psi^\dag G_M^\Pm + U^\Pm G_M^\MM - [1-P_\psi] 
	\left(G_M^\PP U^\Pm + {\rm i} G_V^\Pm \right)
	 \right\}.
	\label{e:eom-for-u-Upm-after-projecting}
	\eeqs
Our goal is small oscillations around the baryon. We write (\ref{e:eom-for-u-Upm-after-projecting}) as a Schrodinger equation, where the wavefunction consists of a vector $u$ and a matrix $U^\Pm$ and the hamiltonian is the pair $(l,L^\Pm)$:
	\beq
	- {\textstyle {\rm i} \frac{{\rm d}}{{\rm d}t}} \pmatrix{ u \cr U^\Pm } 
	= \pmatrix{ l(u,u^\dag, U^\Pm, U^\MP) \cr L^\Pm(u , u^\dag, U^\Pm)}.
	\eeq
However, $(l,L^\Pm)$ depend on $u, U^\Pm$ {\em and} $u^\dag, U^\MP$ through $G_V$ in (\ref{e:eom-for-u-Upm-after-projecting}). Indeed, from \S \ref{a:G-of-V},
	\beq
	G_V^\Pm = 2 {\rm i} G(u \psi^\dag - \psi u^\dag + U^\Pm)^\Pm,  \:\:\:\: 
	{\textstyle \ov{2i}} G_V^\PP = G_{u \psi^\dag - \psi u^\dag }^\PP
	 - G_{U^\MP}^\PP + G_{U^\Pm}^\PP.
	\eeq
So the time dependence does not factorize under separation of variables\footnote{We are looking for vibrations about a time dependent state $\tl \psi_t(p) = \tl \psi_o(p) {\rm e}^{{\rm i}pt/2}$. The momentum-dependent phases in $u$ and $U^\Pm$ guarantee that the gauge conditions $\psi^\dag_t u_t =0$ and $\psi^\dag_t U_t^\Pm =0$ remain satisfied if they initially were.}. This prevented us from finding oscillatory solutions to the full system (\ref{e:eom-for-u-Upm-after-projecting}) using ($\om$ is complex a priori)
	\beq
	\tl u_p(t) = \tl u_p {\rm e}^{ {\rm i} (\om + p/2) t}  \:\:\:\: {\rm and} \;\;\;
	\tl U^\Pm_{pq} = \tl U^\Pm_{pq} {\rm e}^{ {\rm i} \left(\om + (p-q)/2 \right) t}.
	\label{e:separation-of-variables}
	\eeq

\subsection{Eigenvalue problem for oscillations in approximation \texorpdfstring{$u=0$}{} }
\label{s:u-equal-0-approx}


We make an ansatz that permits us to find oscillations around the baryon. $V$ is a meson bound to $M_o$ whose valence-quark wavefunction is $\psi$. $u, U^\Pm$ represent  valence and sea/antiquarks in $V$. Mesons are usually described as a quark-antiquark sea. This suggests putting $u=0$. Moreover, for mesons around the vacuum, $V^\Pm \! \propto \! U^\Pm \! \ne 0$ (\S \ref{s:tHooft-eqn-small-osc}), and our analysis should reduce to that far from the baryon. For $u$ to remain zero under time evolution (\ref{e:eom-for-u-Upm-after-projecting}), a {\tt consistency condition} must hold for $\tl g \ne 0$ 
	\beq
	{\rm i} \dot u = -{\textstyle \frac{\tl g^2}{2}} \left\{ U^\Pm G_M^\MP 
	- {\rm i} (1-P_\psi) G_V^\PP \right\} \psi  \: = \:  0
	 \:\:\:\:\:  {\rm where}  \:\:\:\:  
	G_V^\PP = 2 {\rm i} \left\{ G^\PP_{U^\Pm} - G^\PP_{U^\MP} \right\}.
	\label{e:consistency-cond-on-U-PM}
	\eeq
It says $\psi$ is in the kernel of a certain operator. (\ref{e:consistency-cond-on-U-PM}) is studied in \S \ref{a:solve-consistency-condition}. Hilbert-Schmidt $U^\Pm$ obeying (\ref{e:consistency-cond-on-U-PM}) and $\psi^\dag U^\Pm = 0$ form the {\em physical} subspace for the ansatz $u=0$. Now we assume oscillatory behaviour about the time-dependent g.s. The time-dependence in the eom (\ref{e:eom-for-u-Upm-after-projecting}) factorizes:
	\beq
	\label{e-time-dependence-of-U-PM}
	U^\Pm_{pq}(t) = U^\Pm_{pq} \:\:  {\textstyle {\rm e}^{ {\rm i} \left(\om + \frac{p-q}{2} \right) t }} 
	\; \Rightarrow \;
	\left( \om + {\textstyle \frac{p-q}{2}} \right)  \: U^\Pm_{pq} \:\:  {\rm e}^{ {\rm i} \left( \om + \frac{p-q}{2} \right)t}
	 = L^\Pm(U^\Pm)_{pq} \:\:  {\rm e}^{ {\rm i} \left( \om + \frac{p-q}{2} \right)t}.
	\eeq
Let $K^\Pm(U^\Pm) = L^\Pm(U^\Pm) + [U^\Pm , \frac{{\tt p}}{2}]$. We get an eigenvalue problem for the excitation energies $\om$ above the g.s. of the baryon\footnote{Recall (\ref{e:momentum-of-CHD}) that ${\tt p}$ is the hermitian operator with kernel ${\tt p}_{pq} = 2\pi \delta(p-q) p$.}. The {\em correction} $[U^\Pm , \frac{{\tt p}}{2}]$ accounts for time dependence of the g.s.
	\beq
	K^\Pm(U^\Pm) = 
	[ U^\Pm, {\textstyle \frac{{\tt p} }{ 2}}] 
	+ (1-P_\psi) h U^\Pm - U^\Pm h + {\textstyle \frac{\tl g^2}{2}} \left\{U^\Pm G_M^\MM - (1-P_\psi) \left(G_M^\PP U^\Pm - 2 G_{U^\Pm}^\Pm \right) \right\}  \: = \:  \om U^\Pm.
	\label{e:eigenvalue-problem-for-U-PM}
	\eeq
The {\it eigenvector is a matrix} $U^\Pm$ with $\psi$ in its left nullspace and constrained by (\ref{e:consistency-cond-on-U-PM}). Similarly,
	\beqs
	{\textstyle K^\MP(U^\MP)} \!\!\!\! &=& \!\!\!\!
	[{\textstyle \frac{{\tt p}}{2}},U^\MP]
	+ U^\MP h (1-P_\psi) - h U^\MP + {\textstyle \frac{\tl g^2 }{ 2}} \left\{G_M^\MM U^\MP - \left(U^\MP G_M^\PP - 2 G^\MP_{U^\MP} \right)(1-P_\psi) \right\}  = \om^* U^\MP 
	\nonumber \\ 
	&& \Rightarrow \quad \hat K(U) \; = \; 
	\pmatrix{ 0 & K^\MP(U^\MP) \cr K^\Pm(U^\Pm) & 0 }
	\; = \; \pmatrix{ 0 & \om^* U^\MP \cr \om U^\Pm & 0 }.
	\label{e:eigenvalue-problem-for-U-blocks}
	\eeqs
An advantage of the ansatz $u=0$ is that $K^\Pm$ depends only on $U^\Pm$. $\hat K$ is hermitian with respect to the Hilbert-Schmidt inner-product defined in \S \ref{a:convergence-conditions-inner-product}:
	\beq
	(U, \hat K(\un U)) = (\hat K (U), \un U)
	 \:\:\: {\rm i.e.,} \:\:\: 
	\Re \tr U^\MP \hat K(\un U)^\Pm = \Re \tr \hat K(U)^\MP \un U^\Pm.
	\eeq
Indeed, cyclicity of $\tr$, the gauge condition $U^\MP \psi =0$ and self-adjointness\footnote{This means $\tr U^\MP G^\Pm_{\un U^\Pm} = \tr G^\MP_{U^\MP} \un U^\Pm$, which follows from the definition of $\tl G(U)_{pq}$.}
of $\hat G$ (\ref{e:mat-elts-of-G-hat}) imply
{\setlength\arraycolsep{1pt}
	\beqs \scriptstyle
	\tr U^\MP \hat K(\un U)^\Pm &=&  \scriptstyle
	\tr \left[ U^\MP [\un U^\Pm, \frac{{\tt p}}{2} ]
	+ U^\MP(1-P_\psi)h \un U^\Pm - U^\MP \un U^\Pm h
	+ {\scriptstyle \frac{\tl g^2}{2}} \left\{ U^\MP \un U^\Pm G_M^\MM
	- U^\MP (1-P_\psi) \left( G_M^\PP \un U^\Pm - 2 G^\Pm_{\un U^\Pm}
	\right)	\right\} \right]
	\cr 
	&=& \scriptstyle 
	\tr\left[ [ \frac{{\tt p}}{2} ,U^\MP] \un U^\Pm 
	+ U^\MP h \un U^\Pm -h U^\MP \un U^\Pm
	+ {\scriptstyle \frac{\tl g^2}{2}} \left\{ G_M^\MM U^\MP \un U^\Pm
	- U^\MP G_M^\PP \un U^\Pm + 2 G^\MP_{U^\MP} \un U^\Pm
	\right\} \right] \cr
	&=& \scriptstyle \tr \hat K(U)^\MP \un U^\Pm.
	\label{e:hermiticity-of-linearized-hamiltonian}
	\eeqs}
The original linearized $H(V)$ (\ref{e:time-evol-of-V}) is {\em not} self-adjoint. By  passing from $V \mapsto U$, eliminating redundant variables and imposing $u \!\! = \!\! 0$, we isolated a subspace on which the linearized evolution admits harmonic time-dependance and is formally self-adjoint. $\hat K^\dag \!\! = \!\! \hat K \Rightarrow \om \! = \! \om^*$. The eigenmodes $U^\Pm$ thus describe {\em oscillations} about the baryon. Without translation invariance, we use $P_M = -\tr {\sf p} (M_o + V)/2 = P + P_V$ (\S \ref{a:cons-of-momentum}) as the excitation momentum instead of $P_t$ (\S \ref{s:failure-of-meson-type-ansatz}). So the mass$^2$ per colour is ${\cal M}_{M}^2 = P_M (2 E_M - P_M)$. For small oscillations, 
$E_{M_o+V} \approx E_o + \om$ where $E_o$ is the g.s. energy. $2 E_o \geq P$ where $P$ is the g.s. momentum. In the chiral limit, $2 E_o = P$ (\S \ref{s:regularized-variational-ansatz-baryon-gs}), so 
	\beq
	{\cal M}_{M_o + V}^2 
	= P_M (2 E_M - P_M) 
	\approx (P + P_V) (2E_o + 2 \om - P - P_V)
	\stackrel{m \to 0}{\longrightarrow} (P + P_V) (2 \om - P_V).
	\label{e:mass-of-excited-baryons}
	\eeq
Since $V \ll M_o$, we expect $|P_V| \ll P$, so $P + P_V \approx P > 0$. To ensure\footnote{$2 \om \geq P_t$ for 't Hooft's meson operator (\ref{e:tHooft-meson-eqn}) since meson mass$^2$'s were $\geq 0$ if $m \geq 0$ \cite{thooft-planar-2d-mesons}.} ${\cal M}_{M_o + V}^2 \geq 0$, we need $2 \om \geq P_V - (2E_o -P)$ or in the chiral limit, $2 \om \geq P_V$. But for $u =0$, $P_V =0$ by (\ref{e:mostly-zero-gauge-V-of-U}). So 
	\beq
	u =0  \:\:\: \Rightarrow  \:\:\:\: 
	{\cal M}^2  \: = \:  P(2E_M -P)
	 \: \approx \:  P(2E_o + 2 \om -P)
	 \: \stackrel{m \to 0}{\longrightarrow} \:  2 \om P.
	\label{e:mass-of-excited-baryons-u-zero}
	\eeq
So $\hat K$ and $\om$ should be $\geq 0$ in the chiral limit. Define the parity of {\em meson} $V$ as even if $\tl V_{pq}$ is real-symmetric and odd if it is imaginary-antisymmetric. For the ansatz $u=0$, the eigenvalue equation (\ref{e:eigenvalue-problem-for-U-PM},\ref{e:eigenvalue-problem-for-U-blocks}) follows from a variational principle.  If we extremize ${\cal E} = (U, \hat K(U)) = \tr U^\MP \hat K(U)^\Pm$,
	\beq 
	{\cal E} = \! \tr\left[ \left(h - {\textstyle \frac{{\tt p} }{ 2}} \right) \left\{ U^\Pm U^\MP -  U^\MP U^\Pm \right\}
	+ {\textstyle \frac{\tl g^2 }{ 2}} \left\{ G_M^\MM U^\MP U^\Pm - G_M^\PP U^\Pm U^\MP + 2 G^\MP_{U^\MP} U^\Pm
	\right\}
	\right]
	\label{e:expec-val-of-linearized-egy-K-hat}
	\eeq	
holding $||U||^2 =(U,U) = \tr U^\MP U^\Pm$ fixed via the Lagrange multiplier $\om$, we get (\ref{e:eigenvalue-problem-for-U-PM})
	\beq
	{\textstyle \frac{\delta }{ \delta U^\MP_{qp}}} \left\{ \tr U^\MP_{rs} \hat K(U^\Pm)^\Pm_{sr} - \om \tr U^\MP_{rs} U^\Pm_{sr} \right\} = 0
	 \:\:\:  \Rightarrow   \:\:\: 
	\hat K(U)^\Pm_{pq} = \om U^\Pm_{pq}.
	\label{e:var-ppl-for-eval-problem}
	\eeq
We treated $U^\Pm_{sr}=U^{\MP*}_{rs}$ and $U^\MP_{rs} = U^{\Pm*}_{sr}$ as independent variables and used the fact that $\hat K^\Pm$ depends only on $U^\Pm$. We must solve the eigenvalue problem (\ref{e:eigenvalue-problem-for-U-PM}) on a space of $U^\Pm$ examined in \S \ref{a:solve-consistency-condition}. In \S \ref{s:rank-1-ansatz-for-Upm} we interpret the terms in the variational energy $\cal E$, and approximately minimize it in \S \ref{s:estimate-mass-shape-1st-excitation}.

\subsection{Rank-1 ansatz \texorpdfstring{$U^\Pm = \phi \eta^\dag$}{}: sea quarks and anti-quarks}
\label{s:rank-1-ansatz-for-Upm}

Let $U^\Pm = \phi \eta^\dag$, with $\phi,\eta \in {\cal H}_\pm$ the sea/antiquark wavefunctions of the excited baryon. They have antiquarks even if the lightest one doesn't, just as mesons have antiquarks though the vacuum doesn't. (\ref{e:var-ppl-for-eval-problem}) tells us to hold $\tr U^\Pm U^\MP = ||\phi||^2 ||\eta||^2$ fixed and extremize the linearized energy $(U, \hat K(U))$
	\beq
	{\cal E}(U) = 
	\tr \left(h - {\textstyle \frac{{\tt p} }{ 2}} \right) 
	\left[ ||\eta||^2 \phi \phi^\dag - ||\phi||^2 \eta \eta^\dag \right]
	+ {\textstyle \frac{\tl g^2 }{ 2}} \tr \left[ ||\phi||^2 G^\MM_{M} \eta \eta^\dag 
	- ||\eta||^2 G_M^\PP \phi \phi^\dag + 2 G^\MP_{\eta \phi^\dag} \phi \eta^\dag
	\right]
	\eeq
on the physical subspace. If we factor out $||U||^2 = ||\phi||^2||\eta||^2$ and work with {\em unit vectors} $\phi$ and $\eta$,
	\beq
	{\cal E}(U)/||U||^2  \: = \:  \tr \left[ 
	\left(h- {\textstyle \frac{{\tt p} }{ 2}} \right)	( P_\phi - P_\eta )
	+ {\textstyle {\tl g^2}} \left( G^\MP_{\eta \phi^\dag} \phi \eta^\dag 
	+ {\textstyle \half} P_\eta G_M^\MM 
	- {\textstyle \half} P_\phi G_M^\PP  \right) \right].
	\label{e:variational-energy-sea-partons}
	\eeq
Here $P_\eta = \eta \eta^\dag$ and $P_\phi = {\phi \phi^\dag}$. The variational principle cannot determine $||\phi||$ or $||\eta||$. Recall that $2h = p + {\mu^2 / p}$ with $\mu^2 = m^2 - \tl g^2 /\pi$, so the kinetic and self-energies $\cal T$ of sea-partons is
	\beq
	{\cal T} = \tr 	\left(h- {\textstyle \frac{{\tt p}}{2}} \right)	( P_\phi - P_\eta )
	= {\textstyle \frac{\mu^2}{2}} \int {\textstyle \frac{[{\rm d}p]}{p}} \left[|\tl \phi_p|^2 
	- |\tl \eta_p|^2 \right] .
	\label{e:ke-of-sea-partons}
	\eeq
In the chiral limit ${\cal T} <0$ is purely self-energy. (\ref{e:ke-of-sea-partons}) is valid for excitations around the massless $M_o(t)$ (\ref{e:exp-ansatz-1-parameter-family}). If the lightest baryon were static, then $h- {\textstyle {\tt p}/2} \mapsto h$. Interactions are simply interpreted in position space. As $\phi$, $\eta \in {\cal H}_{\pm}$ the block designations in (\ref{e:variational-energy-sea-partons}) are automatic ($\tr P_\eta G_M^\MM = \tr P_\eta G_M$ etc).
Thus, the Coulomb energy $\tl g^2 {\cal V}_{\rm c}$ of sea quarks $\phi$ interacting with anti-quarks $\eta$ is positive
	\beq
	{\cal V}_{\rm c} = \tr G_{\eta \phi^\dag} \phi \eta^\dag 
		= \int {\rm d}x \:  {\rm d}y \:  |\phi(x)|^2  \: {\textstyle \half} |x-y| \: |\eta(y)|^2 
		= \int {\rm d}x \:  |\phi_x|^2 \:  {\tt v}(x) > 0.
	\label{e:coulomb-energy}
	\eeq
Here ${\tt v}(x) = \half \int |\eta_y|^2 |x-y| \: {\rm d}y$ obeys Poisson's equation.
The exchange interaction of sea-partons and `background' valence-quarks $\psi$ is $\tl g^2 {\cal V}_{\rm e} = \tl g^2 \left( {\cal V}_{\rm e\eta} + {\cal V}_{\rm e\phi} \right)$:
	\beq
	{\cal V}_{\rm e} = {\textstyle \half} \tr \left[ P_\eta G_M - P_\phi G_M \right]
	= \int {\rm d}x \:  {\rm d}y \:  \psi^*(x)\psi(y) \:  {\textstyle \half} |x-y| \: 
	\left\{ \phi(x) \phi^*(y) - \eta(x) \eta^*(y) \right\}.
	\label{e:exchange-energy}
	\eeq
Now $v(x) = {\textstyle \half} \int \psi_y \phi^*_y \:   |x-y| \: {\rm d}y$ and $w(x) = {\textstyle \half} \int \psi_y \eta^*_y |x-y| \: {\rm d}y$ both obey Poisson's equation. Then $V_{\rm e \eta} = \int |w'(x)|^2  \: {\rm d}x > 0$ and $V_{\rm e\phi} = - \int |v'(x)|^2 \: {\rm d}x < 0$. However, $\sgn {\cal V}_{\rm e}$ isn't clear a priori. Thus the energy ${\cal E}= {\cal T} + \tl g^2 ({\cal V}_{\rm c} + {\cal V}_{\rm e})$ has a simple relativistic potential-model meaning. In the chiral limit, the mass of an excited baryon is ${\cal M}^2 = 2 P \om$ where $P$ is the g.s. momentum and $\om = \min {\cal E}$ (\ref{e:mass-of-excited-baryons-u-zero}). 

\subsection{Crude estimate for mass and shape of first excited baryon in chiral limit}
\label{s:estimate-mass-shape-1st-excitation}

\begin{figure}
\centering
\mbox{\subfigure[Level curves of the dimensionless energy ${\tt e}(\alpha,\beta)$.]{\includegraphics[width=6cm]{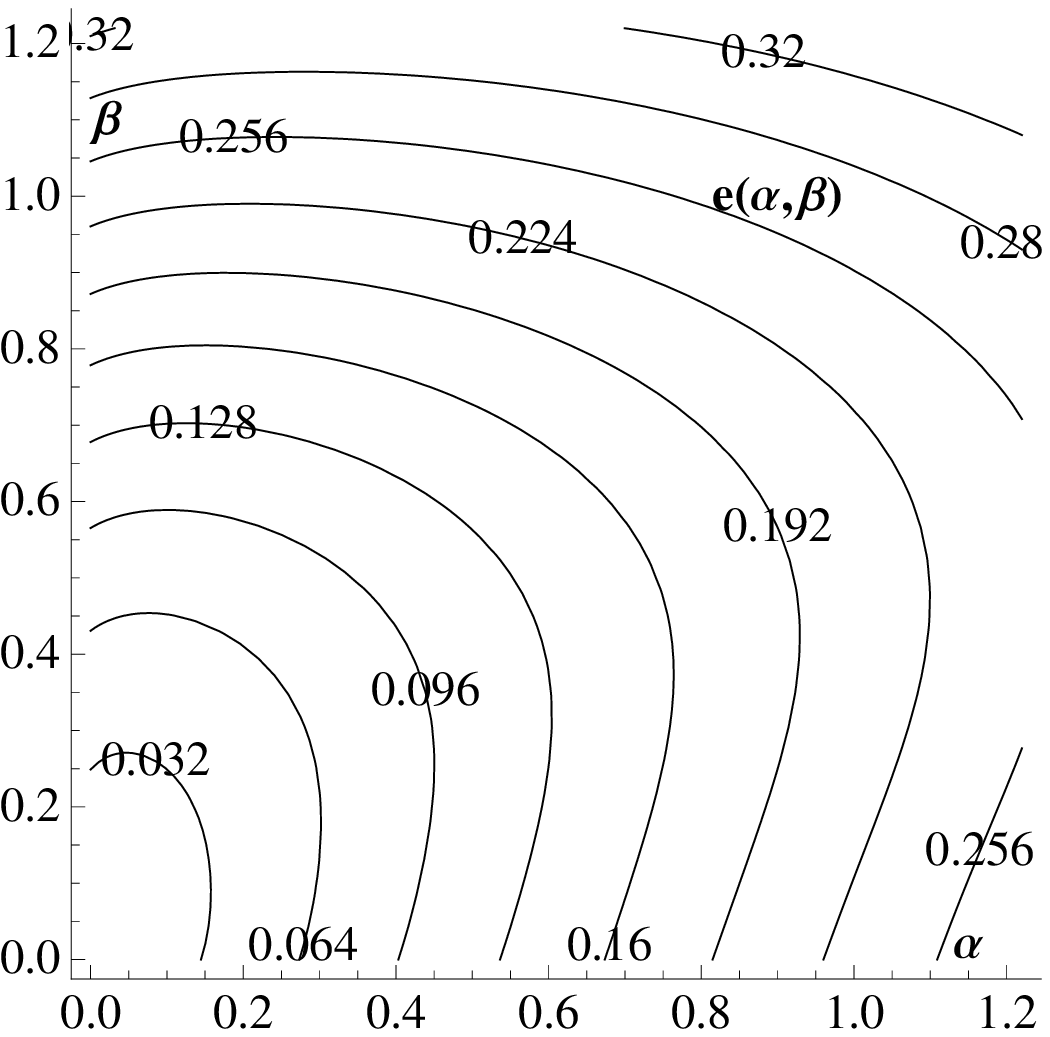}
\label{f:contour-plot-of-e}}
\quad
\subfigure[Valence, sea and anti-quark densities in the excited baryon for $P=1,\alpha=1,\beta=.445$.]{\includegraphics[width=7cm]{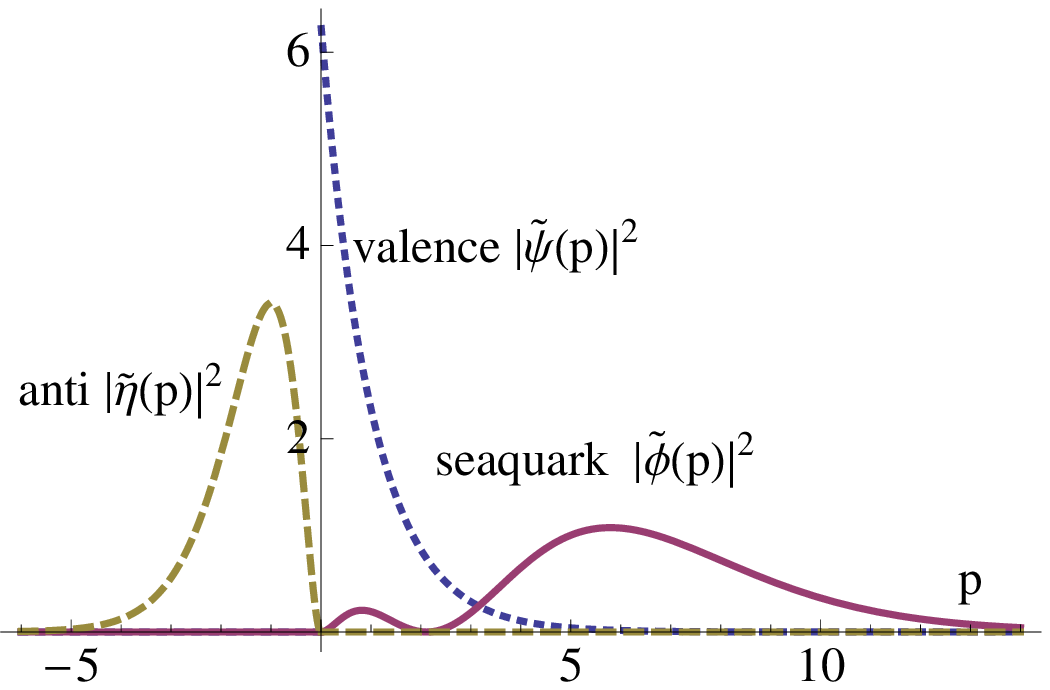} \label{f:val-sea-anti-densities} }}
\caption{\footnotesize (b) The orthogonality of sea and valence ($\phi^\dag \psi =0$ gauge condition) implies that $\tl \phi(p)$ has a node. The normalization of anti/sea distributions is arbitrary, and small compared to the valence distribution. One may contrast these with the first excited meson for which $|\tl \chi(\xi)|^2 \approx \sin^2{\pi \xi}$ where $\xi, 1-\xi$ are the quark and anti-quark momentum fractions.}
\end{figure}

To estimate the mass and form factor $U^\Pm = \phi \eta^\dag$ of the 1st excited baryon (\ref{e:exp-ansatz-for-baryon-gs}), we must extremize ${\cal E}$ (\ref{e:variational-energy-sea-partons}) holding $||U||=1$ and restrict to $U^\Pm$ satisfying the gauge and consistency conditions (\S\ref{a:solve-consistency-condition}). We haven't yet solved the consistency condition (\ref{e:consistency-conds-A-B-for-u-zero}), an intricate orthogonality condition. But even without it, the interacting parton model derived in \S\ref{s:rank-1-ansatz-for-Upm} may be postulated as a mean-field description of excited baryons. So as an approximation, we impose $\psi^\dag \phi =0$ but ignore (\ref{e:consistency-conds-A-B-for-u-zero}). Our ansatz for unit norm $\eta$, $\phi$ contains two parameters $a,b$ controlling the decay of sea parton wavefunctions\footnote{To be accurate in the chiral limit $m \to 0$, $\tl \phi_p$ and $\tl \eta_p$ should probably vanish like small positive powers of $p$ as $p \to 0^\pm$, just as the valence quark wavefunction $\psi$ does. But to keep the calculation of $\cal E$ simple, we chose the smallest integer powers ($\tl \phi_p \sim p^2$ and $\tl \eta_p \sim p$) that ensure absence of IR divergences and orthogonality $\psi^\dag \phi=0$.}
	\beq
	\tl \psi_p = \sqrt{4\pi c} {\rm e}^{-cp} \tht(p), \:\:\: 
	\tl \phi_p = {\textstyle \frac{\sqrt{8\pi b} b^2 (b+c) }{ \sqrt{b^2 + 3c^2}}}
	 	p \left( p- {\textstyle \frac{2 }{ b+c}} \right) {\rm e}^{-bp} \tht(p),  \:\:\: 
	\tl \eta_p = - ap \sqrt{8 \pi a} {\rm e}^{ap} \tht(-p).
	\label{e:variational-ansatz-for-sea-anti}
	\eeq
A boost rescales $p$. We choose our frame by fixing the momentum $P=1/2c$ of the g.s. Since $\tl \phi, \tl \eta$ have been chosen real, $\tl V = {\rm i}[\tl \Phi_o, \tl U] = 2 {\rm i} (0, -\tl \eta \tl \phi^T | \tl \phi \tl \eta^T,0)$ has odd parity, $\tl V^T = -\tl V$. The minimum of ${\cal E} = {\cal T} + \tl g^2 ({\cal V}_{\rm c} + {\cal V}_{\rm e})$ among (\ref{e:variational-ansatz-for-sea-anti}) is the (approx) energy of the $1^{\rm st}$ excited baryon. But it is not an upper-bound, as we ignored (\ref{e:consistency-conds-A-B-for-u-zero}). In the chiral limit, the self-energy is ${\cal T} = {\cal T}_\phi + {\cal T}_\eta$:
	\beq
	{\cal T}_\phi = \tr \left(h- \frac{{\tt p} }{ 2}\right) P_\phi = - \frac{\tl g^2 (3 b^2 - 2 bc + 3c^2) }{ 4 \pi (b^2 + 3 c^2)/b},  \:\:\:\: 
	{\cal T}_\eta = \tr \left(\frac{{\tt p} }{ 2} - h \right) P_\eta = -\frac{\tl g^2 a }{ 2 \pi}.
	\eeq
${\cal T}_\eta, {\cal T}_\phi$ are minimized as $a,b \to \infty$. By real symmetry of $G(M)$ (\S\ref{a:evaluate-G-of-Mo}) and $P_\eta$, the exchange integral
	\beq
	{\cal V}_{\rm e \eta} = {\textstyle \half} \tr P_\eta G_M^\MM = \int [{\rm d}p] \tl \eta_p \int [{\rm d}q] \tl \eta_q G(M)^\MM_{p>q}  = 
	{\textstyle \frac{4 a^2 P }{ \pi (1-2aP)^4}} \left\{ (1-2aP)^2 + 8aP\log{\textstyle \frac{8aP }{ (1+2aP)^2}} \right\}.
	\eeq
${\cal V}_{\rm e \eta} > 0$ since $G(M)^\MM_{pq}$ and $\tl \eta_{q}$ are positive. ${\cal V}_{\rm e \eta}$ increases with $a$, it vanishes at $a=0$. We cross-checked this using ${\cal V}_{\rm e \eta} = \int |w'(x)|^2 {\rm d}x$ (\ref{e:exchange-energy}). $V_{\rm e \phi} = \int {\rm d}x  \: v(x) \:  v^{\pr \pr}(x)^*$ (\ref{e:exchange-energy}) is minimized as $b \to \infty$:
	\beq
	{\cal V}_{\rm e \phi}  \: = \:  - {\textstyle \half} \tr P_\phi G_M^\PP  \: = \:  {\textstyle -\int_0^\infty [{\rm d}p] \:  \tl \phi_p \:  \int_0^p \:  [{\rm d}q] \:  \tl \phi_q \:  G(M)^\PP_{p>q}}  \: = \:  {\textstyle - \frac{2 b^2 P }{ \pi (3 + 4 b^2 P^2)} < 0}.
	\eeq
So the exchange energy is the difference of two +ve quantities $\tl g^2 {\cal V}_{\rm e} = \tl g^2 \left( {\cal V}_{\rm e\eta} + {\cal V}_{\rm e\phi} \right)$. As for the Coulomb energy (\ref{e:coulomb-energy}), ${\cal V}_{\rm c} = \int |\phi(x)|^2 {\tt v} (x)  \: {\rm d}x$, with ${\tt v}(x) = \ov{\pi} \left(a + x \arctan{\frac{x }{ a}} \right)$:
	\beq
	{\cal V}_{\rm c} 
	= {\textstyle \frac{a^2 (a+2 b) \left( b^2+3c^2 \right) + 2 b^2 (2a+b)
   \left(b^2+c^2\right) }{ \pi (a+b)^2 \left(b^2+3 c^2\right)}},
	 \:\:\: {\rm where} \:\:\:  2Pc = 1.
   	\eeq
So ${\cal T}$, ${\cal V}_{\rm e\phi}$ prefer large, while ${\cal V}_{\rm c}$, ${\cal V}_{\rm e\eta}$ prefer small values of $a$ and $b$. What about ${\cal E} = {\cal T} + \tl g^2 \left({\cal V}_{\rm e \phi} + {\cal V}_{\rm e \eta} + {\cal V}_{\rm c} \right)$? $a$ and $b$ are lengths, so define dimensionless parameters $\alpha = a P$ and $\beta = b P$. In the chiral limit the minimum ${\cal M}_1^2$ of $2 {\cal E} P$ is the mass$^2$ of the first excited baryon (\ref{e:mass-of-excited-baryons-u-zero}), so it must be Lorentz-invariant: independent of $P$. $\tl g$ is the only other dimensional quantity, so ${\cal E} = \tl g^2  {\tt e}(\alpha,\beta)/P$, where $\tt e$ is a function of the dimensionless variational parameters. We find
	\beq
	\pi {\tt e} =
	\frac{\alpha}{2}
	- \frac{12 \beta^3 -4\beta^2 +3\beta}{4 \left(4 \beta^2+3 \right)}
	   + \frac{\alpha +2 \beta + 12 \alpha \beta^2 + 8 \beta^3 } {\beta^{-2} (\alpha+\beta )^2 \left(4 \beta^2 + 3\right)}
		- \frac{2 \beta^2}{ 4\beta^2+3 } 
	+ \frac{(1-2 \alpha )^2 + 8 \alpha \log \frac{8 \alpha}{(2 \alpha +1)^2}}{(4 \alpha)^{-2} (1-2 \alpha)^4}
   \eeq
As there is no other scale, the minimum of ${\tt e}$ should be at $\alpha,\beta \sim {\cal O}(1)$. But as plot \ref{f:contour-plot-of-e} of level curves of ${\tt e}$ indicates, the minimum is ${\tt e} =0$ as $\alpha ,\beta \to 0^+$, corresponding to the pathological state where both $\tl \phi, \tl \eta$ (\ref{e:variational-ansatz-for-sea-anti}) tend point-wise to zero! If both $\alpha, \beta$ are free parameters, the minimum occurs on the boundary of the space of rank-1 states $U^\Pm = \phi \eta^\dag$ obeying the gauge condition. Perhaps this was to be expected: without imposing (\ref{e:consistency-conds-A-B-for-u-zero}) we are exploring unphysical states! In the spirit of getting a crude estimate sans imposing (\ref{e:consistency-conds-A-B-for-u-zero}), we put $\alpha = 1$, and minimize in $\beta$ to find $\beta_{\rm min} = .445$ with ${\tt e}(1,\beta_{\rm min}) = .205$. So our crude estimate\footnote{As the plot shows, if we set $\beta =1$ and minimize in $\alpha$, then $\alpha_{\rm min} = .212$ with ${\cal M}=.32 \tl g$, which is roughly the same.} for the mass/colour of the 1st excited baryon in the chiral limit is ${\cal M}_1 = .29 \tl g$. Plot \ref{f:val-sea-anti-densities} has the approximate valence, sea and antiquark densities (\ref{e:variational-ansatz-for-sea-anti}) with parameters $aP = 1$, $bP = \beta_{\rm min}$ and $2cP = 1$. The momentum/colour $P$ of the lightest baryon sets the frame of reference. However, this is not an upper-bound on the mass gap, ${\cal M}_1$ could be an under-estimate as we did not impose (\ref{e:consistency-conds-A-B-for-u-zero}). There is still the unlikely possibility of zero modes other than the 1-parameter family of states associated with the motion of the lightest baryon (\S \ref{s:lightest-baryon}).

\section{Discussion}
\label{s:discussion}
\begin{figure}
\centerline{\includegraphics[width=4.8in]{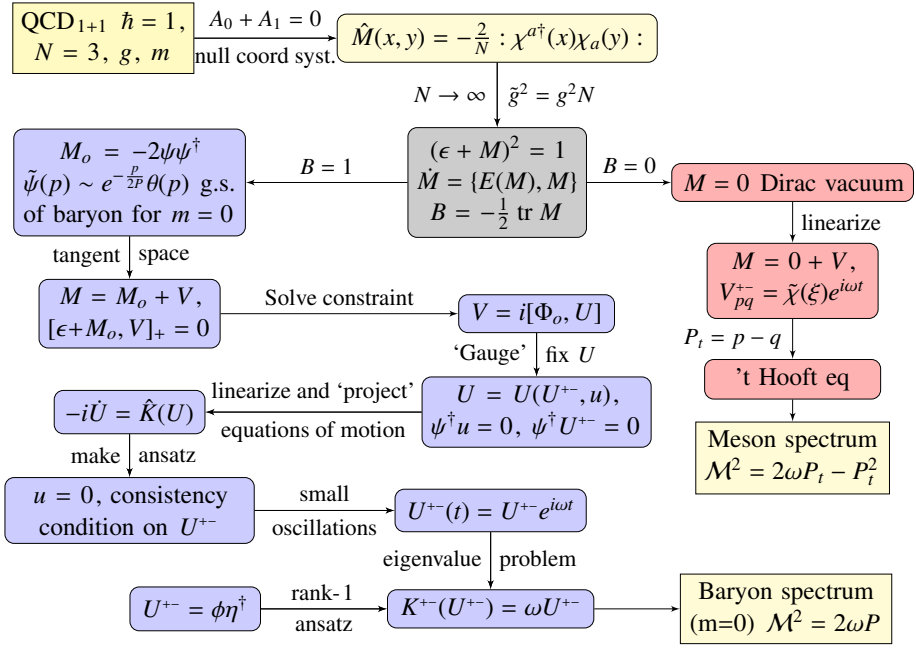}}
\caption{\footnotesize Flowchart of our approach to the baryon spectrum of large-$N$ QCD$_{1+1}$.}
\label{f:flowchart}
\end{figure}

We found that the lightest baryon has zero mass/colour in the chiral limit of large-$N$ QCD$_{1+1}$. There is no spontaneous chiral symmetry breaking in this sense. Being massless, it evolves at the speed of light into a family of massless even parity states (\S \ref{s:gs-in-meson-sector}). They have the same quark distributions $\tl M(p,p)$, differing only in off-diagonal form factors $\tl M_0(p,q) {\rm e}^{{\rm i}(p-q)t/2}$. The other {\em modulus} of the baryon is its size $1/P$. $P$ is its mean momentum/colour, fixed by the frame. Excited baryons (small oscillations around $M_o$) are like bound states of a meson $V$ with $M_o$. Upon eliminating redundant variables we derived an approximate eigenvalue problem for a singular integral operator to determine form factors $U^\Pm$ and masses of excited baryons\footnote{However, we haven't quite solved the consistency condition for the approximation $u=0$ (\S \ref{a:solve-consistency-condition}) which restricts the space of physical states $U^\Pm$. It is also of interest to find a way of proceeding without this approximation.}. Based on the ansatz $U^\Pm = \phi \eta^\dag$ we derived an interacting mean-field parton model for the structure of excited baryons (\S \ref{s:rank-1-ansatz-for-Upm}). Using simple trial anti/seaquark wavefunctions $\eta, \phi$, we estimated the mass and shape of the first excited baryon for which $V$ has odd parity (analogue of Roper resonance). The baryon $M_o$ breaks translation invariance, deforms the vacuum and consequently deforms the shape of the meson $V$. Unlike mesons $\tl \chi(\xi)$ near the Dirac vacuum where $\xi \leftrightarrow 1-\xi$ relates quark and anti-quark distributions, the distribution of quarks $|\tl \phi_p|^2$ and anti-quarks $|\tl \eta_p|^2$ in $V$ aren't simply related. By linearising around $M_o$, we approximated these excited baryons as non-interacting and stable. The non-linear/linear treatment of $M_o/V$ also prevented us from assigning a parity to excited baryons. But their non-linear time evolution (\ref{e:hamiltons-eqns}) should contain information on interactions and decay. Our approach is summarized in figure \ref{f:flowchart}.

{\noindent \bf Acknowledgements:} We thank the UK EPSRC for a fellowship and the referees for their questions.

\appendix

\section{Conservation of mean momentum \texorpdfstring{$P_M = - \half \tr {\tt p} M$}{}}
\label{a:cons-of-momentum}

$E_M$ and $P_M$ were used to define the mass (\ref{e:mass-squared-defn}) of the baryon and of oscillations above a non-translation-invariant $\tl M_t(p,q)$, where the other concept of momentum $P_t = p-q$ is not meaningful (see \S \ref{s:tHooft-eqn-small-osc}). Here we show $P_M = -\half \int p \tilde M_{pp} [{\rm d}p]$ is conserved even if $M(x,y;t)$ is not static, as long as it decays sufficiently fast: $|M_{xy}|^2 \sim {|x|^{-1-\delta}}$ for some $\delta > 0$ as $|x| \to \infty$ for each $y,t$. When $\tl g =0$, energy $T = -\half \tr hM$ is linear. Also, ${\tt \tl p},\tl h$ and $\tl \eps$ are diagonal, so their commutators vanish. From (\ref{e:pb-of-linear-fns}),
	\beq
	\pdr_t P = \{ T(M), P \} = \{ f_h, f_{\tt p} \} = f_{-{\rm i}[h, {\tt p}]} + {\textstyle \frac{{\rm i}}{2}} \tr [h, {\tt p}] \eps = 0.
	\eeq
So for $g \ne 0$ only $U$ (\ref{e:potn-egy-U-intermsof-GofM}) contributes to $\pdr_t P_M$. $U$ is simpler in position-space, so write
	\beq {\textstyle
	P_M = -{\textstyle \half} \int [{\rm d}p] \: {\rm d}x \: {\rm d}y  \: p \:  {\rm e}^{- {\rm i} p (x-y)} M_{xy} 
	= -{\textstyle \half} \int {\rm d}x \:  {\rm d}y \:  M_{xy} \:  D_{xy}}
	\eeq
where $D_{xy} = \int [{\rm d}p] p {\rm e}^{- {\rm i} p (x-y)} = {\rm i} \pdr_x \delta(x-y)$ is hermitian. So we have a quadruple integral
	\beq \textstyle
	\pdr_t{P}  \: = \:  \{ E(M), P \}  \: = \:  \{U, P \} 
	 \: = \:  -{\textstyle \frac{\tl g^2}{16}} \int {\rm d}x \:  {\rm d}y \:  {\rm d}z \:  {\rm d}u \:  {\textstyle \frac{|x-y| }{ 2}} D_{zu} \:  \{ M_{xy} M_{yx} , M_{zu} \}.
	\eeq
We do two integrals and integrate by parts elsewhere to show $\pdr_t P =0$! By (\ref{e:pb-of-M}), the P.B. is
	\beq 
	 {\rm i} \{ M_{xy} M_{yx}  \: , \:  M_{zu} \}  \: = \:  \delta_{yz} \:  M_{yx} \:  \Phi_{xu} \:  
	- \:  \delta_{xu}  \: M_{yx} \:  \Phi_{zy}  \: + \:  (x \leftrightarrow y). 
	\eeq
After one integration and relabelling variables, $\pdr_t {P} = -{\textstyle \frac{\tl g^2 }{ 8}} \,  \Im \, I$, where $I  = {\rm i} \int {\rm d}y \,  {\rm d}z \, \Phi_{yz} \, \int {\rm d}x \, |x-y| \,  M_{xy} \, \pdr_x \delta_{xz}$.
Integrate by parts on $x$ noting that the boundary term $B_1(y,z) = \left[ \: |x-y| \delta_{xz} M_{xy} \:  \right]^\infty_{-\infty} =0$,
	\beq \textstyle
	I = - {\rm i} \int {\rm d}y \:  {\rm d}z \:  \Phi_{yz} \:  |z-y| \:  \pdr_z M_{zy}
	- {\rm i} \int {\rm d}y \:  {\rm d}z \:  \Phi_{yz} \:  \sgn(z-y) \:  M_{zy}.
	\eeq
The 2nd term is real and does not contribute to $\Im I$. So
	\beq \textstyle
	\pdr_t {P}  \: = \:  {\textstyle \frac{\tl g^2 }{ 8}} \Re \int {\rm d}x \:  {\rm d}y \:  \Phi(y,x) \:  |x-y|  \: \pdr_x M(x,y)   \: \equiv \:  {\textstyle \frac{\tl g^2}{8}} \Re J.
	\eeq
Integrating by parts, the boundary term vanishes if $M$ falls off sufficiently fast\footnote{From (\ref{s:review-chd}) $\eps_{yx} \sim {\rm i}  (\pi x)^{-1}$ as $|x| \to \infty$ for any fixed $y$. So the $1{\rm st}$ term in $B_2$ vanishes if $M_{xy} \to 0$ as $|x| \to \infty$.
	\beq \textstyle
	B_2  \: = \:  \int {\rm d}y  \: \left[ \:  \left\{ \eps_{yx} + M_{yx} \right\} \:\:  |x-y| M_{xy}  \: \right]^\infty_{-\infty}
	\label{e:boundary-term-B2}
	\eeq
The $2{\rm nd}$ term in $B_2$ vanishes iff $\lim_{|x| \to \infty} |M_{xy}|^2 |x-y| = 0$, for any fixed $y$. This $2{\rm nd}$ condition subsumes the first. So $B_2=0$ provided $|M_{xy}|^2 \sim |x|^{-1 - \delta}$ for some $\delta > 0$. This is easily satisfied by our ansatz $M_o(x,y)$ (\ref{e:exp-ansatz-for-baryon-gs}) for the baryon g.s.}
	\beq \textstyle
	J = B_2 - \int {\rm d}x {\rm d}y \:  M_{xy} \:  \Phi_{yx} \sgn(x-y)
	- \int {\rm d}x {\rm d}y  \: M_{xy}  \: |x-y|  \: \pdr_x \eps_{yx} 
	- \int {\rm d}x {\rm d}y  \: M_{xy}  \: |x-y|  \: \pdr_x M_{yx}.
	\eeq
The first two integrals are imaginary and do not contribute to $\Re J$, so
	\beq \textstyle
	\pdr_t {P} = - {\textstyle \ov{8}} \tl g^2 \:  \Re K,  \:\:\:\: 
	{\rm where \:\: } K  \: = \:  \int {\rm d}x \: {\rm d}y \:  M_{xy}  \: |x-y| \:  \pdr_x M_{yx}.
	\eeq
Integrating by parts we express $L = K + K^* = 2 \Re K = - \int {\rm d}x \: {\rm d}y \:  |M_{xy}|^2 \sgn(x-y) + B_3$. $B_3 = \int {\rm d}y  \: [ \:  |M_{xy}|^2 |x-y|  \: ]^\infty_{-\infty}$ is familiar from $B_2$ (\ref{e:boundary-term-B2}), and vanishes under the same hypothesis. Finally, $\sgn$ is odd, so $\pdr_t {P} = - \tl g^2 L/16 =0$. So $P_M$ is conserved if $|M_{xy}|^2$ decays as ${x^{-1-\delta}}$ for some $\delta > 0$.

\section{Finite part integrals (Hadamard's {\em partie finie})}
\label{a:finite-part-integrals}

A finite part integral is like an ODE: rules to integrate the singular measure are like boundary conditions (b.c.). Here we define the $1/p^2$ singular integrals appearing in the potential energy. In position-space this is manifested in the linearly rising $|x-y|$ potential. 't Hooft \cite{thooft-planar-2d-mesons} defines them by averaging over contours that go above/below the singularity. Here we formulate them via real integrals and physically motivate \& justify the definition by showing it satisfies the relevant b.c. Both methods use analytic continuation. Consider the rank-$1$ baryon \S \ref{s:mass-of-exp-ansatz} and suppose ${\rm support} \:  \tl \psi \subseteq [0,P]$,
    \beq \textstyle
   PE  = \, \int [{\rm d}p] \, \tl \psi(p) \: \finttxt [{\rm d}s] \, \tl \psi^*(p+s) \, \tl V(s), \; {\rm where} \;\; \tilde V(s)= - s^{-2} \tl W(s).
    \eeq
Recall that $V'' = |\psi|^2$ with two b.c. (\ref{e:bc-for-self-cons-potn-in-posn-space}). So $\tilde V(s) = -s^{-2} \int [{\rm d}q] \tl \psi(s+q) \tl \psi^{*}(q)$ is singular at $s=0$. Here $\tl W^*(s) = \tl W(-s)$, i.e., $\Re \tl W(s)$ is even and $\Im \tl W(s)$ is odd\footnote{From (\S\ref{s:regularized-variational-ansatz-baryon-gs}), if $\tl \psi(p)$ is (dis)continuous at $p=0$, then so is $\tl W'(s)$ at $s=0$. If $\tl \psi(p) \sim p^a$, then $\tl W(s) -1 \sim |s|^{1+ 2a}$.}. Now, the two b.c. imply
    \beqs \textstyle
    V(0) = -\finttxt  \Re \tl W(s) \frac{[{\rm d}s] }{ s^2} =
        \int |\psi(y)|^2 \frac{|y| }{  2} {\rm d}y,  \:\:\: 
    V'(0) = \finttxt  \Im \tl W(s) \frac{[{\rm d}s] }{  s} =
    	- \half \int {\rm d}y \:  |\psi(y)|^2 \sgn y.
        \label{e:boundary-cond-for-poisson-eqn}
    \eeqs
The lhs of (\ref{e:boundary-cond-for-poisson-eqn}) don't exist as Riemann integrals since $\tl W(0)=1$. But the rhs exist quite often and can be used to define the lhs. E.g. rhs of $V(0)$ makes sense if $\psi$ decays faster than $1/y$. The rhs of $V'(0)$ makes sense as long as $\psi(y)$  decays faster than $|y|^{-\half}$. This includes $|\psi(y)| \sim 1/|y|$ as $|y| \to \infty$ corresponding to $\tl \psi(p)$ having a jump discontinuity. In particular, it can be used to define $\finttxt {\tl W(s)} s^{-1}  {\rm d}s$ even when $\tl W'(p)$ is discontinuous at $p =0$. Now we eliminate $\psi$ and express singular integrals of $W$ in terms of Riemann integrals of $W$. For simplicity, suppose $\tilde \psi(p) \in \mathbf{R}$. Then $\psi(-x) = \psi^*(x)$, and $\tl W$ is real and even. The $V'(0)$ b.c. (\ref{e:boundary-cond-for-poisson-eqn}) is satisfied. Let us also restrict attention to wavefunctions such that $\tilde \psi(p) \sim p^a$, $a>0$ as $p\to 0$. Our aim is to define $ -\finttxt \frac{1}{s^2} \tl W(s) [{\rm d}s]$ so as to satisfy the first b.c. The rule should reduce to the Riemann integral, when this quantity is finite to begin with.

{\noindent \bf Claim:} Let $\tl W(s)$ be even and $\tl W'(0) = 0$ 
For $P> 0$, if we define
    \beq \textstyle
        \finttxt_{-P}^{P} \frac{1}{s^2} \tl W(s) [{\rm d}s] :=
        \int_{-P}^{P} \frac{\tl W(s) - \tl W(0) }{ s^2} [{\rm d}s] -
        \frac{\tl W(0) }{ \pi P}, \:\:\: 
	{\rm then} \:\:\:\:\: 
        \finttxt_{-P}^{P} \frac{1}{s^2} \tl W(s) [{\rm d}s] = -
        \int_{-\infty}^{\infty} |\psi(x)|^2 \frac{|x|}{2} {\rm d}x.
    \label{e:defn-of-finite-part-integral}
    \eeq
{\noindent \bf Proof:} We subtracted divergent terms and analytically continued what we'd have got if $\tl W(s)$ vanished sufficiently fast at the origin (i.e. $W(s) \sim s^{1+\eps},  \:  \eps > 0$) to make the integral converge. The main point is that this definition satisfies the $V(0)$ b.c. (\ref{e:boundary-cond-for-poisson-eqn}). Recall that $W$ is the charge density:
    \beq \textstyle
    \tl W(s) = \int_{-\infty}^{\infty} |\psi(x)|^2 {\rm e}^{-{\rm i}sx} {\rm d}x,
	 \:\:\: {\rm so \:\:  that} \:\:\: 
	\tl W(s) - \tl W(0) = \int_{-\infty}^{\infty}
        	{\rm d}x |\psi(x)|^2 ({\rm e}^{-{\rm i}sx} - 1).
    \eeq
Moreover, $\tl W'(0) = - {\rm i} \int_{-\infty}^{\infty} x |\psi(x)|^2 {\rm d}x =0$ as the integrand is odd. Therefore, $\tl W(s) - \tl W(0)$ vanishes at least as fast as $s^{1+\eps},  \:  \eps > 0$ as $s \to 0$. E.g. for $\tilde \psi(p) \propto p^a {\rm e}^{-p}, \:  \tl W(s) -1 \propto - s^{2a + 1} + O(s^2)$. Therefore, $\int_{-P}^P \{ \tl W(s) - \tl W(0) \}  s^{-2} [{\rm d}s] < \infty$. As the integrand is even it suffices to consider
    \beq \textstyle
        \int_{0}^P \frac{\tl W(s) - \tl W(0) }{ s^2} [{\rm d}s] = \int_0^P \frac{{\rm d}s}{
        2\pi s^2} \int_{-\infty}^{\infty} {\rm d}x |\psi(x)|^2 ({\rm e}^{-{\rm i}sx} -1).
    \eeq
Only the even part of $({\rm e}^{-{\rm i}sx} - 1)$ contributes to the integral on $x$. Reversing the integrals,
    \beq \textstyle
        \int_0^P \frac{\tl W(s) - \tl W(0) }{ s^2} [{\rm d}s]
        = \:  \int_{-\infty}^{\infty} {\rm d}x |\psi(x)|^2 \left( \ov{2\pi P} - \nu(x) \right).
    \eeq
This involves the sine integral, $2\pi P \nu(x) = Px \:  {\rm Si}(Px) + \cos(Px)$.
Now $\tl W(0) =1$, so
    \beq \textstyle
        \int_0^P \frac{\tl W(s) - \tl W(0) }{ s^2} [{\rm d}s] - \frac{\tl W(0) }{ 2\pi P}
        = -\int_{-\infty}^{\infty} {\rm d}x |\psi(x)|^2 \nu(x).
    \eeq
We must show $\nu(x)$ may be replaced by $|x|/4$ under the integral. Since Si$(t)$ is odd, we have
    \beq \textstyle
    \nu(x) = \frac{|x| }{ 4} + \frac{1 }{ 2\pi P}
        	\left( Px \:  {\rm Si}(Px) - \frac{P|x|\pi }{ 2} + \cos(Px) \right)
        = \frac{|x| }{ 4} + \frac{R(Px) }{ 2\pi P},
    \eeq
where $R(t) = t \:  {\rm Si}(t) - |t| \pi/2 + \cos{t}$. We have the desired result except for a remainder term:
    \beq \textstyle
        \int_{-P}^{P} \frac{\tl W(s) - \tl W(0) }{ s^2} [{\rm d}s] - \frac{\tl W(0) }{
        \pi P} = - \int_{-\infty}^{\infty} |\psi(x)|^2 \frac{|x|}{2} {\rm d}x - \frac{1}{\pi P} \int_{-\infty}^{\infty} |\psi(x)|^2  \:  R(Px) {\rm d}x.
    \eeq
When $P \to \infty$, the remainder term $\to 0$ as $|R(t)| \leq 1$. For finite $P$, $R(t) \sim \frac{-\sin{t}}{t}, |t| \to \infty$ is oscillatory\footnote{The asymptotic expansion of ${\rm Si}(t)$ for large $t$ is ${\rm Si}(t) \sim \frac{\pi}{2} + \left(- \ov{t}	+ {\cal O}(t^{-3}) \right) \cos{t} + \left(-\ov{t^2} + {\cal O}(t^{-4}) \right) \sin{t}$.}, so we expect the remainder term to be small. But it is zero. Consider
    \beqs \textstyle
    \int_{-\infty}^{\infty} {\rm d}x |\psi(x)|^2  \:  R(Px)
    	\!\!\! &=& \!\! \textstyle \int_0^P [{\rm d}q] \int_{-q}^{P-q} [{\rm d}r]
        \tl \psi(q+r) \tl \psi^*(q) \int_{-\infty}^{\infty} {\rm d}x {\rm e}^{ {\rm i} r x}  \: R(Px). \cr
	R(t) {\rm  \:\: is \:\:  even \:\:  and} \!\!\! && \!\!\!\!\!
        \int_{-\infty}^{\infty} {\rm d}x {\rm e}^{ {\rm i} r x}  \:  R(Px) = 2
        \int_{0}^{\infty} {\rm d}x \cos(rx)  \:  R(Px) = 0,
    \eeqs
from the properties of ${\rm Si}$, provided $|r| < P$, which is the region of interest. Thus the remainder term vanishes, and we have shown that our definition of the ``finite part'' integral satisfies the b.c. This justifies our definition (\ref{e:defn-of-finite-part-integral}) when $\tl W(s)$ is even and $\tilde W'(0) =0$. q.e.d.

According to (\ref{e:defn-of-finite-part-integral}), $\finttxt_{-P}^P \frac{{\rm d}r}{r^2} = -\frac{2 }{ P}$. Moreover, it makes sense to define $\finttxt_{-P}^P \frac{{\rm d}r}{r} :=0$ since the integrand is odd. We use these to  extend the definition to functions on an even interval $[-P,P]$ but with $W'(0)$ possibly non-zero. Suppose $W(s)$ is continuously differentiable at $s=0$ with $W(s)-W(0)-s W'(0) \sim s^{1+\eps}$ for some $\eps >0$ and $s$ sufficiently small. Then we define
	\beq \textstyle
	\finttxt_{-P}^P \frac{{\rm d}s }{ s^2} W(s) := 
	\int_{-P}^P \frac{{\rm d}s }{ s^2} \left[ W(s) - W(0) - s W'(0) \right] 
	- \frac{2 }{ P} W(0).
	\label{e:def-fin-part-extended}
 	\eeq
This is used to evaluate $\hat G(M_o)$ in \S\ref{a:evaluate-G-of-Mo}. In general, this rule is applied in a small neighbourhood $[-\eps,\eps]$ of the singularity. The 1$^{\rm st}$ term on the rhs of (\ref{e:def-fin-part-extended}) vanishes as $\eps \to 0$ giving
	\beq \textstyle
	\finttxt_{-\infty}^\infty W(s) \frac{{\rm d}s }{ s^2} := \lim_{\eps \to 0} \left[ \left\{ \int_{-\infty}^{-\eps} + \int_\eps^\infty \right\} W(s) \frac{{\rm d}s }{ s^2} - \frac{2 }{ \eps} W(0) \right].
	\eeq

\section{Interaction operator \texorpdfstring{$\hat G$}{} and \texorpdfstring{$\hat G(M)$}{} for baryonic vacua}
\label{a:evaluate-G-of-Mo}

$\hat G$ is the operator on hermitian $M$ defining (\ref{e:potn-egy-U-intermsof-GofM}) the potential energy $8 U = \tl g^2 \tr M \hat G(M)$. $\hat G(M)$ is a hermitian matrix with kernel $G(M)_{xy} = \half M_{xy} |x-y|$ or $\tl G(M)_{pq} = -\finttxt \frac{[{\rm d}r]}{r^2} \tl M_{p+r,q+r}$. The null-space of $\hat G$ consists of diagonal $M_{xy} = m(x) \delta(x-y)$, which don't lie on the phase space (\ref{e:Gr-1}) except for $M = 0$. $U$ is positive definite. The matrix elements of $\hat G$ are real
	\beq \textstyle
	\hat G_{xy}^{zw} = {\textstyle \half} {|x-y|} \delta(x-z) \delta(w-y)
	 \:\:\:\: {\rm where} \:\:\:  
	G(M)_{xy} = \int {\rm d}z \: dw \:  \hat G_{xy}^{zw} M_{zw}.
	\label{e:mat-elts-of-G-hat}
	\eeq
The entries $\hat G^{zw}_{xy}$ are symmetric under a left-right flip $\hat G_{xy}^{zw} = \hat G_{yx}^{wz}$, which means $M \mapsto G(M)$ preserves hermiticity. Moreover $\hat G_{xy}^{zw} = \hat G_{wz}^{yx}$, which implies $\hat G$ is hermitian as an operator on hermitian matrices (\S\ref{a:self-adj-of-operator-on-herm-mat}). In momentum space, $\tl G^{rs}_{pq} = \tl G^{sr}_{qp} = \tl G^{pq}_{rs}$ are real, with $\tl G(M)_{pq} = \int [{\rm d}r \:  {\rm d}s] \:  \tl G^{rs}_{pq} \: \tl M_{rs}$. Here $\tl G_{pq}^{rs} = -\finttxt \frac{[{\rm d}t] }{ t^2}  \:  \delta^r_{p+t} \delta_{q+t}^s$ and $\delta_p^q \equiv 2\pi \delta(p-q)$. $G(M)_{xy}$ is simple, but the Fourier transform $\tl G(M)_{pq}$ is sometimes more convenient to solve the eom (e.g. \S \ref{s:degeneracy-time-evol-baryon-gs},(\ref{e:eigenvalue-problem-for-U-PM})). At the baryon vacua $M(\tau)$ (\ref{e:exp-ansatz-1-parameter-family}):
	\beq
	\tl G(M(\tau))_{pq}  \: = \:  - {\rm e}^{ \frac{ {\rm i} }{ 2} (p-q) \tau } \finteq {\textstyle \frac{[{\rm d}r] }{ r^2}} \tl M(0)_{p+r,q+r}
	 \: = \:  {\rm e}^{ \frac{ {\rm i} }{ 2} (p-q) \tau } G(M_o)_{pq}.
	\label{e:G-of-M-of-tau}
	\eeq
So it suffices to take $\tau=0$. For $M_o = -2 \psi_o \psi_o^\dag$  (\ref{e:exp-ansatz-for-baryon-gs}) with $\tl \psi_o$ real, $\tl G_{M_o}$ is symmetric. $G(M)_{xy}$ isn't rank-$1$. But $\psi_o(p+r) \sim {\rm e}^{-p} {\rm e}^{-r} \tht(p+r)$ factorizes, ensuring that $G(M_o)^{\pm \mp}$ are rank-1. In general,
	\beq
	\tl G(M_o)_{pq} = \frac{2 }{ P} \exp{ \left( - \frac{{p+q} }{ 2 P} \right)} \finteq_{\max(-p,-q)}^\infty \frac{{\rm d}r}{r^2} {\rm e}^{-r/P}.
	\label{e:integral-for-GofMo}
	\eeq
If $p \; {\rm or} \;  q<0$, then $t \equiv \max(-p,-q) = -\min(p,q)> 0$ and there is no singularity:
	\beq {\textstyle
	{\rm I}_2(t)  \: = \:  \int_t^\infty \frac{{\rm d}r }{ r^2} {\rm e}^{-r/P}
		 \: = \:  \frac{{\rm e}^{-t/P} }{ t}
	+ \ov{P} {\rm Ei}\left(- \frac{t }{ P} \right)  \: > \:  0,  \:\:\:\:  {\rm for} \:\:\:\:  t > 0.}
	\label{e:I2}
	\eeq
Here $\textrm{Ei}(z) = - \int_{-z}^\infty \frac{{\rm e}^{-u} }{ u} {\rm d}u$. $I_2(t)$ monotonically decays from $\infty$ to $0$ exponentially, as $t$ goes from $0$ to $\infty$. Thus, in the $(p,q) = (-+), (+-)$ and $(--)$ quadrants,
	\beq
	\tl G(M_o)_{pq} = {\textstyle \frac{2 }{ P} \exp{ \left( - \frac{p+q}{2 P} \right)} \left( \frac{{\rm e}^{-t/P} }{ t} + \ov{P} {\rm Ei}\left(-\frac{t }{ P} \right) \right)},
	 \:\:\:  {\rm where} \:\:\:  t = -\min(p,q) > 0.
	\label{GofM-PM-MP-MM-blocks}
	\eeq
In the $++$ quadrant, $s = \min(p,q) > 0$ so we may write
	\beq {\textstyle
	\tl G(M_o)^\PP_{pq} = -\ov{2\pi} \tl M_o(p,q) \:  {\rm I}(s)
	 \:\:\:\:  {\rm where} \:\:\:  {\rm I}(s)
	= \left[ \finttxt_{-s}^{s} + \int_{s}^\infty \right] \frac{{\rm d}r}{r^2} {\rm e}^{-r/P} = {\rm I}_1 + {\rm I}_2. }
	\eeq
Here ${\rm I}_1(s)$ is a finite part integral defined in (\ref{e:def-fin-part-extended}), and expressed via the sinh integral
	\beq {\textstyle
	{\rm I}_1(s) = \finttxt_{-s}^{s} \frac{{\rm d}r}{r^2} {\rm e}^{-r/ P}
		:= -\frac{2}{s}
	+ \int_{-s}^{s} \frac{{\rm d}r}{r^2} \:  \left\{{\rm e}^{-r/ P} -1 + \frac{r }{ P} \right\}
	= -\frac{2}{s} \cosh\left(\frac{s}{P} \right) 
	+ \frac{2}{P} \:  {\rm Shi}\left(\frac{s}{P} \right).}
	\eeq
Here, ${\rm Shi}(z) = \int_0^z \frac{\sinh(t) }{ t} {\rm d}t$. Combining with the previously encountered ${\rm I}_2(s)$ (\ref{e:I2}),
	\beq {\textstyle 
	{\rm I}(s) =  {\rm I}_1 + {\rm I}_2 = - \ov{s} {\rm e}^{s/P} + \frac{2 }{ P} \:  {\rm Shi}\left(\frac{s }{ P} \right)
	+ \ov{P} \textrm{Ei}\left(-\frac{s }{ P}\right)
	= - \ov{s} {\rm e}^{s/ P} + \ov{P} \left({\rm Chi}\left(\frac{s }{ P} \right) + {\rm Shi}\left(\frac{s }{ P} \right) \right).}
	\eeq
${\rm Chi}(z) = \gamma + \log z + \int_0^z \frac{\cosh{t} -1 }{ t} {\rm d}t$. Now we summarize $\tl G(M_o)_{pq}$ in all blocks. Let $s = \min(p,q)$, then
	\beqs
	\tl G(M_o)_{pq} &=& {\textstyle \frac{2 }{ P} \exp{ \left( -\frac{p+q}{2P} \right)}
	\left\{ \begin{array}{ll}
	I_2(-s) = -\ov{s} {\rm e}^{s/P} + \ov{P} \textrm{Ei}(\frac{s}{P}) & \textrm{if $s < 0$} \\
	I(s) = -\ov{s} {\rm e}^{s/P} + \ov{P} \left( \textrm{Chi}(\frac{s}{P}) + \textrm{Shi}(\frac{s}{P}) \right) & \textrm{if $s > 0$}.
	\end{array} \right.}	\cr
	&=& {\textstyle \frac{2}{P} \exp{ \left( -\frac{p+q}{2 P} \right)}
	\pmatrix{ I_2\left(-\min(p,q)\right) & I_2\left(-p \right)
	\cr I_2 \left(-q \right) & I\left(\min(p,q)\right) }
	}.
	\label{e:G-of-Mo-summary}
	\eeqs
$I_2(t)$ monotonically decays from $\infty$ to $0$ exponentially, as $t$ goes from $0$ to $\infty$. $I(s)$ monotonically grows from $-\infty$ to $\infty$ for $0 < s < \infty$. The factor $\frac{2}{P} \exp{ \left( -\frac{p+q}{2P} \right)}$ $=$ $-\ov{2\pi} \tl M_o(p,q)$, but only for $p,q>0$. $G(M_o)$ inherits some properties of $M_o$: $G(M_o)^\Pm_{pq} = f(p) g(q)$ is rank-$1$ like $M_o$ and $V^\MP M_o^\PP =0$ implies that $V^\MP G(M_o)^\Pm =0$ (\S \ref{s:H2--zero-consistency-condition}). But $G(M_o)$ {\em doesn't} commute with $M_o$, $\eps$ or $\Phi_o$.

What if $s=\min(p,q)=0$, which is the boundary of the $++$ quadrant? From (\ref{e:integral-for-GofMo}), when $s = 0$, $\tl G(M_o)_{pq} \propto \finttxt_0^\infty \frac{{\rm d}t}{t^2} {\rm e}^{-t}$, which cannot be prescribed a finite value\footnote{Recall $\tl G(M_o)_{pq} = \half \int {\rm d}x {\rm d}y M_o(x,y) |x-y| {\rm e}^{- {\rm i} (px-qy)}$. For $s=0$, an oscillatory phase is absent. As $M_o(x,y)|x-y| \sim x^0$, the integral diverges. The divergence is absent on a space of finite length or for $M(x,y)$ decaying faster at infinity.}. $\tl G(M_o)_{pq}$ is continuous everywhere except along $s = 0$. It approaches $\pm\infty$ as $s \to 0^\pm$. However, its derivative is discontinuous across the line $p=q$. It decays exponentially to zero in all directions except along the positive $p$ or $q$ axes.

\subsection{Interaction operator \texorpdfstring{$\hat G(V)$}{} in terms of \texorpdfstring{$U$}{}}
\label{a:G-of-V}

Since $V^\MM =0$ (\ref{e:constraint-linear-blockform}) for a tangent to the phase space at the lightest baryon $M_o(t)$, there are some simplifications in $G_V(t)$. Let $s= \max(p,q)$, then $G(V)_{pq} = -\finttxt_{-s}^\infty \frac{[{\rm d}r]}{r^2} \tl V_{p+r,q+r}$. Due to the positive support of $\tl M_{pq}$, $G_V^\MM$ never appears in the eom. 
In the mostly-zero gauge (\ref{e:mostly-zero-gauge-V-of-U})
	\beq
	\tl G(V)_{p >q} 
	 = 2 {\rm i} \tl G\left(u \psi^\dag - \psi u^\dag + U^\Pm \right)_{p>q}, \quad
	\tl G(V)_{p <q} 
	= 2 {\rm i} \tl G\left(u \psi^\dag - \psi u^\dag - U^\MP \right)_{p<q}.
	\eeq
Of course, $u,\psi, V,U$ are all time-dependent. In particular, if $u=0$ as in \S \ref{s:u-equal-0-approx}, we write compactly
	\beq
	G^\Pm_V = 2 {\rm i} G^\Pm_{U^\Pm}, \:\:\:\: 
	G^\MP_V = -2 {\rm i} G^\MP_{U^\MP} 
	 \:\: {\rm and} \:\:  
	G_V^\PP = 2 {\rm i} \left\{ G^\PP_{U^\Pm} - G^\PP_{U^\MP} \right\} = 2 {\rm i} G^\PP_{U^\Pm} + h.c.
	\eeq

\section{Completing proof that \texorpdfstring{$M(t)$}{} solves equations of motion}
\label{a:Z-equals0}

In \S \ref{s:degeneracy-time-evol-baryon-gs} we studied the time evolution of the baryon states $M(t)$ (\ref{e:exp-ansatz-1-parameter-family}). In the chiral limit, the eom is $	{\textstyle \frac{ {\rm i} }{ 2}} \dot M_{pq} = {\textstyle \ov{4}} M_{pq} \left( q-p \right) + \frac{\tl g^2 }{ 4} Z(M(t))_{pq}$ (\ref{e:eom-chiral-limit}). We show here that the interaction terms $\propto Z(t)$ identically vanish for our massless states $M(t)$ ($Z$ stands for zero). Recall that
	\beq {\textstyle
 	Z(t)_{pq}
	 \: = \:  \ov{\pi} \left(\ov{p} - \ov{q} \right) \tl M_t(p,q) - G(M_t)_{pq} \{\sgn p - \sgn q \} + [G(M_t), M_t]_{pq} =  Z_1 + Z_2 + Z_3.}
	\eeq
It is seen that $Z(t)_{pq} = Z(0)_{pq} \exp{[ \frac{ {\rm i} }{ 2} (p-q) t ]}$. We show here that $Z_{pq} \equiv Z(0)_{pq} = 0$. Now $M_o$ and $G(M_o)$ (\S \ref{a:evaluate-G-of-Mo}) are real-symmetric, so $Z_{1,2,3}(p,q)$ are real-antisymmetric. $Z_1$ is simplest
	\beq
	Z_1(p,s) = \pi { \textstyle \left(\ov{p} - \ov{s} \right)} \tl M_{ps} = \frac{4 }{ P} {\rm e}^{-(p+s)/2P} {\textstyle \left( \ov{s} - \ov{p} \right)} \tht(p) \tht(s).
	\label{e:z1plusplus}
	\eeq
$Z_2(p,s) = -G(M)_{ps} \{\sgn p - \sgn s \} $ vanishes in the $ps = ++, --$ quadrants while
	\beq
	(Z_2)^\Pm_{ps} = -2 G(M)_{ps}^\Pm  \:\:\:\: {\rm and } \:\:\:\:  (Z_2)^\MP_{ps} = 2 G(M)_{ps}^\MP.
	\eeq
Since $\tl M_o$ has positive support, $Z_3^\MM = [G(M_o),M_o]^\MM =0$. So $Z^\MM = 0$. What about the other quadrants? To proceed, we need $G(M_o)_{pq}$, from (\ref{GofM-PM-MP-MM-blocks}). In the $-+, --, +-$ quadrants,
	\beq
	G(M_o)_{pq} = \frac{2 }{ P} {\rm e}^\frac{-(p+q) }{ 2P}
	\left\{ \begin{array}{ll}
	-\ov{p} {\rm e}^{p/P} + \ov{P} \textrm{Ei}(p/P) & \textrm{if $p<0, p<q$} \\
	-\ov{q} {\rm e}^{q/P} + \ov{P} \textrm{Ei}(q/P) & \textrm{if $q <0, q < p$.}
	\end{array} \right.
	\eeq
This is enough to evaluate $Z_2^\Pm$, (anti-symmetry determines $Z_2^\MP$, while $Z_2^\PP = Z_2^\MM =0$)
	\beq {\textstyle
	Z_2^\Pm(p,s) = \frac{4 }{ P} {\rm e}^{- (p+s)/ 2P} \bigg( \ov{s} {\rm e}^{s/P}
	- \ov{P} \textrm{Ei}\left( \frac{s}{P} \right) \bigg)
	{\rm  \:\:\:\: for \:\:\: } p > 0 > s.}
	\eeq
This is also adequate to find $Z_3^\Pm$ and $Z_3^\MP$. For example,
	\beq {\textstyle
	Z_3^\Pm(p,s) = -\int_0^\infty [{\rm d}q] \tl M_{pq}^\PP G(M_o)_{qs}^\Pm  \: = \: -\frac{4}{P} {\rm e}^{-(p+s)/2P} \bigg( \ov{s} {\rm e}^{s/P}
	- \ov{P} \textrm{Ei}\left(\frac{s}{P} \right) \bigg).}
	\eeq
We see that $Z_2^\Pm + Z_3^\Pm = 0$. As $Z_1^\Pm =0$ we conclude $Z^\Pm = 0$. By antisymmetry, $Z^\MP=0$. 

{\noindent \sf ++ Block:} Here $Z^\PP = Z_1^\PP + Z_3^\PP$ with $Z_3^\PP = [G(M_o)^\PP, M_o^\PP]$. For $Z_3^\PP$ we need $G(M_o)_{pq}^\PP = \frac{2}{P} {\rm e}^{-(p+q)/2P}  \: {\rm I}[\min(p,q)]$ (\ref{e:G-of-Mo-summary}). Antisymmetry allows us to consider $0 < p \leq s$,
	\beqs
	{\textstyle Z_3^\PP(p,s)} &=& {\textstyle \frac{4}{P^2} {\rm e}^{-\frac{p+s}{2P}}
		\int_0^\infty {\rm d}q \:  {\rm e}^{-\frac{q}{P}}
		\bigg\{ {\rm I}[\min(q,s)] - {\rm I}[\min(p,q)] \bigg\}} \cr
	&=& {\textstyle \frac{4}{P^2} {\rm e}^{-\frac{p+s}{2P}} \bigg[ P \bigg\{ {\rm I}(s) {\rm e}^{-\frac{s}{P}} - {\rm I}(p) {\rm e}^{-\frac{p}{P}} \bigg\} + \int_p^s {\rm d}q \:  {\rm e}^{-\frac{q}{P}} {\rm I}(q)
	\bigg].}
	\eeqs
This is anti-symmetric in $p, s$, so it is valid for all $p,s>0$. The integral is expressed as
	\beqs
	&& {\textstyle \int_p^s {\rm d}q \:  {\rm e}^{-q/P} {\rm I}(q)  \: = \:  
	{\rm e}^{-{p / P}} {\rm Ei}\left(\frac{p}{P} \right)- {\rm e}^{-{s / P}}
   {\rm Ei}\left(\frac{s}{P}\right),  \:\:\:\: {\rm so} }
	\cr
	{\textstyle Z_3^\PP(p,s) \!\!\!\!} &=& {\textstyle \!\!\!\! 
	\frac{4}{P^2} {\rm e}^{-\frac{p+s}{2P}}
		\bigg[{\rm e}^{-\frac{p}{P}} \bigg\{ \textrm{Ei}\left(\frac{p}{P} \right)
		- P  \: \textrm{I}(p) \bigg\} -
		(s \leftrightarrow p) \bigg]
	= \frac{4}{P} {\rm e}^{-\frac{p+s}{2P}} \left(\ov{p} - \ov{s} \right).}
	\label{e:Z3plusplus}
	\eeqs
From (\ref{e:z1plusplus},\ref{e:Z3plusplus}), $Z^\PP = Z_1^\PP + Z_3^\PP = 0$. So $Z(t) \equiv 0$ and $M(t)$ (\ref{e:exp-ansatz-1-parameter-family}) solves the chiral eom.

\section{Convergence conditions and inner product on perturbations}
\label{a:convergence-conditions-inner-product}

The phase space of QCD$_{1+1}^{N=\infty}$ is the Grassmannian Gr$_1$ (\ref{e:Gr-1},\cite{rajeev-2dqhd}). To define an integer-valued baryon number labelling components of $Gr_1$, we need the convergence condition $\tr [\eps ,M]^\dag[\eps, M] < \infty$, i.e., $[\eps, M]$ is Hilbert-Schmidt. 
Applying this to $M = M_o + V$ the condition on a tangent vector $V$ is
	\beq
	2 \tr [\eps, V]^\dag [\eps, M_o] + \tr |[\eps, V]|^2 < \infty.
	\label{e:conv-cond-perturbation}
	\eeq
The 1st term is $0$ for the g.s. $M_o = -2 \psi \psi^\dag$ with $\eps \psi = \psi$ since $[\eps, M_o]=0$. Decomposing $V$ in blocks (\ref{e:constraint-linear-blockform}), (\ref{e:conv-cond-perturbation}) becomes $\tr V^\Pm V^\MP < \infty$, i.e., $V^\Pm$ is H-S. Also, $\tr V^\PP < \infty$ must be trace class (\S 4.1 of \cite{rajeev-2dqhd}). There is a natural positive-definite symmetric inner product $(V,\un V) = \tr V \un V$ on the tangent space to Gr$_1$ if we further assume that $V^\MM, V^\PP$ are H-S. We use it to define self-adjointness of the hamiltonian for linearized evolution in (\ref{e:hermiticity-of-linearized-hamiltonian}). At the baryon g.s. $M_o = -2\psi \psi^\dag$, $V^\MM=0$, so writing $V = {\rm i} [\Phi_o,U]$ and expressing $U$ in the mostly-zero gauge (\ref{e:mostly-zero-gauge-V-of-U}), the inner product is
	\beq
	(V, \un V) = \tr V \un V
	= 2 \Re \tr V^\MP \un V^\Pm + \tr V^\PP \un V^\PP
	= 4 (U, \un U)
	= 8 \Re \tr \left( U^{\MP} \un U^{\Pm} + u \un u^\dag \right).
	\eeq

\section{Hermiticity of a linear operator on hermitian matrices}
\label{a:self-adj-of-operator-on-herm-mat}

A transformation $U \mapsto K(U)$ on hermitian matrices must preserve hermiticity. If  $K(U)_{pq} = \hat K_{pq}^{rs} U_{rs}$, this becomes $\left(\hat K^{rs}_{pq} - \hat K^{sr*}_{qp} \right) U_{rs} = 0$ $\forall$ hermitian $U$. We can't conclude $\hat K^{rs}_{pq} = \hat K^{sr*}_{qp}$, this isn't necessary as $U_{rs} = U_{sr}^*$ aren't independent. We go to a basis for hermitian matrices
    \beq 
    [R_{ab}]_{pq} = \delta_{ap} \delta_{bq} + \delta_{aq} \delta_{bp}, \;\;
    [I_{ab}]_{pq} = {\rm i} \left(\delta_{ap} \delta_{bq} - \delta_{aq} \delta_{bp} \right),
    \label{e:basis-for-hermitian-matrices}
    \eeq
and deduce the necessary and sufficient conditions\footnote{Here $K^{[rs]}_? = K^{rs}_? + K^{sr}_?$ and $K^{\{rs\}}_? = K^{rs}_? - K^{sr}_?$ while $?$ is held fixed.} for $\hat K$ to preserve hermiticity of $U$
	\beq
	\hat K^{[rs]}_{pq} = \hat K^{[rs]*}_{qp} \quad {\rm and} \quad
	\hat K^{\{rs\}}_{pq} = -\hat K^{\{ rs \}*}_{qp}.
	\eeq
What does it mean for such a $\hat K$ to be formally self-adjoint? The space of hermitian matrices has the inner-product $(U, U') = \tr U U'$. So self/skew-adjointness is the condition
	\beq
	(\hat K U , U') = \pm (U, \hat K U') \;\; {\rm or} \;\;
	\tr K(U) U' = \pm \tr U K(U') \;\;\; \forall \;\;\; U,U' \;\; {\rm  hermitian}.
	\eeq
So $\forall$ hermitian $U,U'$: $\hat K^{qp}_{sr} \:  U_{qp} \:  U'_{rs} = \pm \:  \hat K^{rs}_{pq} \:  U_{qp} \:  U'_{rs}$. A sufficient condition for $\hat K$ to be self/skew-adjoint is (anti-)symmetry under left-right {\em and} up-down flips of indices: $\hat K^{qp}_{sr} = \pm \:  \hat K^{rs}_{pq}$. Using (\ref{e:basis-for-hermitian-matrices}), necessary and sufficient conditions for self/skew-adjointness of $\hat K$ are
    \beq
    \hat K^{[ab]}_{[cd]} = \pm \hat K^{[cd]}_{[ab]}, \quad
    \hat K^{\{ab\}}_{\{cd\}} = \pm \hat K^{\{cd\}}_{\{ab\}} \quad
    {\rm and } \quad
    \hat K^{\{ab\}}_{[cd]} = \mp \hat K^{[cd]}_{\{ab\}}.
    \label{e:condition-for-self-adj}
    \eeq

\section{Space of physical states consistent with \texorpdfstring{$u=0$}{} ansatz}
\label{a:solve-consistency-condition}

The physically motivated (\S \ref{s:u-equal-0-approx}) ansatz $u=0$ led to a hermitian eigenvalue problem for the baryon spectrum (\ref{e:eigenvalue-problem-for-U-PM}). We imposed it so that the equation for perturbations around the g.s. (\ref{e:eom-for-u-Upm-after-projecting}) admits oscillatory solutions via variable separation, by removing simultaneous dependence on both $U^\Pm$ and $U^\MP$. $U^\Pm: {\cal H}_- \to {\cal H}_+$ must be H-S (\S \ref{a:convergence-conditions-inner-product}) and respect the gauge $\psi^\dag U^\Pm = 0$ and consistency condition (\ref{e:consistency-cond-on-U-PM}) for $u(t)$ to remain $0$. Here we examine (\ref{e:consistency-cond-on-U-PM}). Momentum-dependent phases (\ref{e-time-dependence-of-U-PM}) cancel, leaving
	\beq
	{\rm e}^{ {\rm i} \om t} U^\Pm G_M^\MP \psi + 2(1 - P_\psi) \left( {\rm e}^{ {\rm i} \om t} G^\PP_{U^\Pm} \psi - {\rm e}^{- {\rm i} \om t} G^\PP_{U^\MP} \psi \right) =0.
	\eeq
So the coefficients of ${\rm e}^{\pm {\rm i} \om t}$ must vanish, leaving two time-independent vector conditions
	\beq
	{\rm (A):}  \:\:\:  \left\{ U^\Pm G_M^\MP + 2 (1^\PP - P_\psi) G^\PP_{U^\Pm} \right\} \psi =0
	 \:\:\:\: {\rm and} \:\:\:\: 
	{\rm (B):}  \:\:\:  (1^\PP - P_\psi) G^\PP_{U^\MP} \psi = 0,
	\label{e:consistency-conds-A-B-for-u-zero}
	\eeq
on a whole operator $U^\Pm$. We expect a large space of solutions $U^\Pm$. (\ref{e:consistency-conds-A-B-for-u-zero}) says $\psi$ is annihilated by a pair of operators built from $U^\Pm$: another type of orthogonality between ground/excited states. (B) is simpler than (A). Introducing an arbitrary $n \in {\cal H}_-$ and $\la \in \mathbf{C}$,
	\beq
	(B): \:\:\:\:  (1^\PP - P_\psi) G^\PP_{U^\MP} \psi = 0   \:\: \Leftrightarrow \:\:  G_{U^\MP} \psi = \la \psi + n.
	\eeq
Let us look for rank-$1$ solutions $U^\Pm = \phi \eta^\dag$ with $\phi, \eta \in {\cal H}_\pm$ the sea and anti-quark wavefunctions of the meson $V$ bound to the baryon $M_o$. We solve for $\phi^*(x) = \ov{\psi} \left(\frac{ \la \psi + n }{\eta} \right)^{\pr\pr}$.
For $\phi$ to lie in ${\cal H}_+$, $\phi^*(x)$ must necessarily be analytic in $\mathbf{C}^-$\footnote{A necessary (but not sufficient) condition for $\tl \psi(p)$ to be a positive momentum function ($\psi \in {\cal H}_+$), is for $\psi(x)$ to be the boundary value of a function holomorphic in the upper half of the complex $x$ plane $\mathbf{C}^+$.}. We argue this requires $\la =0$. $\psi(x) \propto (c-ix)^{-1}$ doesn't have zeros (\ref{e:exp-ansatz-1-parameter-family}), but it has a pole in $\mathbf{C}^-$, which can't be cancelled by either $\eta(x)$ or $n(x)$, both of which are analytic in $\mathbf{C}^-$. Thus $\la=0$, and in particular $G(\eta \phi^\dag)^\PP \psi = 0$: an interaction operator built from $U$ annihilates the g.s. So rank-1 solutions of (B) are of the form $\phi^*(x) = \ov{\psi} (n/\eta)^{\pr \pr}$, parameterized by vectors $n, \eta \in {\cal H}_-$\footnote{We haven't {\em proved} that $\phi \in {\cal H}_+$. There may be more conditions on $n,\eta$ to {\em guarantee} $\phi \in {\cal H}_+$.}. For e.g., $\eta = (a+ {\rm i} x)^{-2} \in {\cal H}_-$, $n = (a + {\rm i}x)^{-m} \in {\cal H}_-$, $m > 2$ and $\phi \propto (2Px -{\rm i})(a - {\rm i} x)^{-m} \in {\cal H}_+$ is a family of solutions of (B) with $P, a > 0$. 

We haven't yet solved (A) in such generality. Here we give a restricted class of solutions of (A), where each term of (A) is zero. For $U^\Pm = \phi \eta^\dag$ we get two conditions on $\phi$ and $\eta$
	\beq
	{\rm (A1)} \:\:\:\:\:  \phi \:  \left(\eta^\dag G_M^\MP \psi \right) =0
	 \:\:\:\:\:  {\rm and}  \:\:\:\: 
	{\rm (A2)} \:\:\:\:\:  (1^\PP - P_\psi) G^\PP_{\phi \eta^\dag} \psi =0.
	\eeq
(A1) $\Rightarrow  \eta^\dag G_M^\MP \psi =0$: the antiquark wavefunction must be $\perp$ to $G_M^\MP \psi$\footnote{If $\psi_o$ is the baryon g.s. (\ref{e:exp-ansatz-for-baryon-gs}), $\eta$ must be $\perp$ to $(G_M^\MP \psi)_{p < 0} = \sqrt{\frac{2}{\pi P}} {\rm e}^{-\frac{p}{2P}} \left\{- \frac{{\rm e}^{\frac{p}{P}}}{p} + \ov{P} {\rm Ei}(\frac{p}{P}) \right\}$.  $(G_M^\MP \psi)_p$ is +ve and exponentially decays monotonically from $\infty$ to $0$ as $p$ goes from $0$ to $-\infty$. $(G_M^\MP \psi)_p \sim -\ov{p} \sqrt{2/\pi P}$ as $p \to 0^-$. To avoid IR divergences, $\tl \eta(p) \sim (-p)^\gamma$ for some $\gamma >0$ as $p \to 0^-$.}. For $P=1$, $\tl \eta_p = p(p+.474) {\rm e}^p \tht(-p)$ is such a function. (A2) $\Leftrightarrow G_{\phi \eta^\dag} \psi = \la^\pr \psi + m$ for arbitrary $\la^\pr \in \mathbf{C}$ and $m \in {\cal H}_-$. (A2) resembles (B), but they aren't the same though ${G_{U^\MP}}^\dag = G_{U^\Pm}$. We solve (A2) for $\eta^*(x) = \ov{\psi} \left( \frac{\la^\pr \psi + m }{ \phi}  \right)^{\pr \pr}$. As before, there are conditions for this $\eta$ to lie in ${\cal H}_-$.
But it is possible that (A1) \& (A2) form too small a class of solutions of (A). We haven't yet combined (A) \& (B) to find $U^\Pm$ obeying (\ref{e:consistency-conds-A-B-for-u-zero}). We hope to remedy this in the future.

\footnotesize




\end{document}